\documentclass[
 reprint,
showpacs,preprintnumbers,
 amsmath,amssymb,
 aps,
prb,
]{revtex4-1}

\usepackage{graphicx}
\usepackage{dcolumn}
\usepackage{bm}
\usepackage{mathrsfs}

\usepackage{color}
\usepackage{ulem}
\usepackage{comment}

\allowdisplaybreaks[1]
\newcommand{\average}[1]{\ensuremath{\langle#1\rangle} }

\begin{document}

\title{Extrinsic spin Hall effect in inhomogeneous systems}
\author{Takumi Funato}
\author{Hiroshi Kohno}
\affiliation{Department of Physics, Nagoya University, Nagoya 464-8602, Japan}
\date{\today}

\begin{abstract}
 Charge-to-spin conversion in inhomogeneous systems is studied theoretically.  
 We consider free electrons subject to impurities with spin-orbit interaction 
and with spatially modulated distribution, 
and calculate spin accumulation and spin current induced by an external electric field. 
 It is found that the spin accumulation is induced by the vorticity of electron flow  
through the side-jump and skew-scattering processes, 
and the differences of the two processes are discussed.  
 The results can be put in a form of generalized spin diffusion equation with a spin source term 
given by the divergence of the spin Hall current.
 This spin source term reduces to the form of spin-vorticity coupling 
when the spin Hall angle is independent of impurity concentration. 
We also study the effects of long-range Coulomb interaction, 
which is indispensable when the process involves charge inhomogeneity and accumulation 
as in the present problem. 
\end{abstract}

\maketitle

\section{introduction}
Interconversion between charge and spin currents is one of the fundamental processes in spintronics.
Among various methods, the direct and inverse spin Hall effects (SHE) via spin-orbit interaction (SOI) 
have been utilized most commonly
\cite{d'yakonov_perel, hirsch1, zhang1, murakami1, sinova1, engel1, kato1, wunderlich1,  saitoh1, valenzuela1, zhao1, she_rev}.
 Recently, there are several experimental reports on the enhancement 
of charge-to-spin conversion in naturally oxidized (nox-) Cu \cite{ando1,nozaki1,nitta}.
 An {\it et al.} found an enhanced spin-orbit torque generation efficiency comparable to Pt in nox-Cu 
in the spin-torque ferromagnetic resonance (ST-FMR)\cite{sot_rev} experiment\cite{ando1}. 
 Okano {\it et al.} measured charge-spin interconversion in nox-Cu by unidirectional spin Hall magnetoresistance (USMR)\cite{avci1, avci2, usmr_theo} and spin pumping\cite{mizukami1, urban1, sp_rev},
  and demonstrated its non-reciprocal character\cite{nozaki1}. 
 A possibly related phenomenon was reported by Enoki {\it et al.}, 
who observed enhanced weak antilocalization in nox-Cu\cite{nitta}.
 There are also reports on the enhancement of spin-orbit torques in nox-Co 
by Hibino {\it et al.}\cite{hibino} and Hasegawa {\it et al.}\cite{hasegawa}

Previous studies suggested that the spin-vorticity coupling (SVC)\cite{matsuo4, shd_theory, basu2} 
can be the main origin of the enhanced charge-to-spin conversion in nox-Cu. 
 The SVC is the coupling between the electron spin and the vorticity of electron flow that arises 
in the rotating ({\it i.e.,} non-inertial) frame of reference locally fixed to the lattice or the electrons.
 This effect has been demonstrated by several experiments, 
which include spin hydrodynamic generation in liquid metals \cite{shd} 
and spin-current generation from surface acoustic waves in solids.\cite{nozaki_saw} 
 However, applying the idea of SVC to the aforementioned phenomena in nox-Cu 
seems to contain a difficulty in the treatment of moving lattice in the non-inertial frame 
comoving with the electrons.

 The purpose of this paper is to study charge-to-spin conversion phenomena that occur 
in inhomogeneous systems which allow electron flow with vorticity. 
 We model the system by nearly free electrons placed under impurities with SOI and 
with spatially modulated distribution. 
 We calculate spin accumulation and spin current in response to an applied electric field 
using Kubo formula\cite{kubo} and discuss their relation to the vorticity of the electron flow. 
 We found a spin accumulation is induced by the vorticity of the electron flow 
via the side-jump and skew-scattering processes.  
 The obtained results are precisely described by a generalized spin diffusion equation 
with a spin source term given by the divergence of the spin Hall (SH) current. 
 The spin source term reduces to the vorticity of the electron flow if the SH angle 
is independent of the impurity concentration, which occurs for skew scattering in the present model.
 This may be considered as an ``effective SVC" viewed in the laboratory (inertial) frame.

 In the course of the study, we noticed that the processes necessarily involve 
electron charge inhomogeneity in the equilibrium state as well as in the nonequilibrium states. 
 To treat such situations properly, we also study the effects of long-range Coulomb interaction.

 This paper is organized as follows. 
 In Sec.~II, we describe the (basic) model as well as its extension to include long-range Coulomb interaction. 
 The calculation for the model without Coulomb interaction is outlined in Sec.~III, 
and the results are summarized in Sec.~IV, 
where a generalized spin diffusion equation is derived and applied to a thin film system. 
 We study the effects of long-range Coulomb interaction in Sec.~V. 
 The conclusion is given in Sec.~VI. 
 Details of the calculation are presented in Appendices. 
 In Appendix A, the ladder vertex correction is given. 
 The response functions are calculated in Appendix B (without Coulomb interaction) and 
in Appendix C (with Coulomb interaction). 
 Some discussion on the screening of SOI is given in Appendix D.  
 Some integrals are given in Appendix E.

 Throughout this paper, we set $\hbar =1$.

\section{model}

 We consider a normal metal in which impurities, which cause normal and spin-orbit scattering, 
are distributed with slight and spatially slow inhomogeneity. 
 We define the \lq\lq basic model'' that does not consider the long-range Coulomb interaction 
in Sec.~\ref{basic model}, and introduce the Coulomb interaction in Sec.~\ref{model with Coulomb}.

\subsection{Basic model}
\label{basic model}

 We consider a nearly free electron system in two dimensions (2D) or three dimensions (3D) 
in the presence of SOI due to impurities.  
The Hamiltonian is given by
\begin{align}
H= \sum_{\bm k} \epsilon_{\bm k}
\psi^{\dagger}_{\bm k}\psi_{\bm k} + H_{\text{imp}} + H_{\text{so}}, 
\label{H}
\end{align}
where $\epsilon_{\bm k} = \frac{k^2}{2m}$ is the kinetic energy, and 
$\psi^{\dagger}_{\bm k} = (\psi_{{\bm k} \uparrow}^\dagger , \psi_{{\bm k} \downarrow}^\dagger)$ 
and $\psi_{\bm k}$ are the electron creation (annihilation) operators.
$H_{\text{imp}}$ and $H_{\text{so}}$ describe the coupling to the impurity potential and impurity SOI, respectively, 
\begin{align}
 & H_{\text{imp}} =\sum_{\bm k, \bm k'} V_{\bm k-\bm k'} 
   \psi^{\dagger}_{\bm k} \psi_{\bm k'} , 
\label{imp}
 \\
 & H_{\text{so}} = i\lambda_{\text{so}}\sum_{\bm k, \bm k'} 
  V_{\bm k-\bm k'}(\bm k \times \bm k') \cdot 
   \psi^{\dagger}_{\bm k} \bm \sigma \psi_{\bm k'},
\label{soi}
\end{align}
where $\lambda_{\text{so}}$ is the strength of SOI, and 
$\bm \sigma = (\sigma^x, \sigma^y, \sigma^z)$ are Pauli matrices. 
 We assume a short-range impurity potential, 
$V_{\text{imp}}(\bm r) = u_{\text i} \sum_j \delta (\bm r-\bm R_j)$, 
where $u_{\text i}$ is the strength of the potential and $\bm R_j$ is the position of $j$th impurity, 
and $V_{\bm k - \bm k'}$ is the Fourier transform of  $V_{\text{imp}}(\bm r)$.

 As a model of inhomogeneous systems, we consider a situation 
in which the impurity distribution has a slight spatial modulation.
 Specifically, we take 
$n_{\text i}(\bm r) = n_{\text i} + \delta n_{\text i} (\bm Q) \, e^{i\bm Q\cdot \bm r}$ 
for the impurity concentration, where the first term is the ``uniform'' part and the second term 
is the ``inhomogeneous'' part with $\bm Q$ being the wave vector of the modulation. 
 Using the probability density function of impurities,
\begin{align}
 p_{\text{imp}} ({\bm r}) 
= \frac{1}{\Omega} + \frac{\delta n_{\text i} (\bm Q)}{n_{\text i}}\frac{e^{i\bm Q\cdot \bm r}}{\Omega},
\label{eq:p_imp}
\end{align}
where $\Omega$ is the system volume, 
we average over the impurity positions as $\left< V_{\bm k}\right>_{\text{av}}=\delta n_{\text i} (\bm Q) u_{\text i}\delta_{\bm k, \bm Q}$,
$\left< V_{\bm k}V_{\bm k'}\right>_{\text{av}}=n_{\text i}u_{\text i}^2\delta_{\bm k+\bm k', 0}+\delta n_{\text i} (\bm Q) u_{\text i}^2\delta_{\bm k+\bm k', \bm Q}$, and
$\left< V_{\bm k}V_{\bm k'}V_{\bm k''}\right>_{\text{av}}=n_{\text i}u_{\text i}^3\delta_{\bm k+\bm k'+\bm k'', 0}+\delta n_{\text i} (\bm Q) u_{\text i}^3\delta_{\bm k+\bm k'+\bm k'', \bm Q}$. 
 In this paper, we consider  the inhomogeneity of the impurities ($\delta n_{\text i} (\bm Q)$) 
up to the first order, and the impurity SOI ($\lambda_{\text{so}}$) up to the second order.
 The impurity-averaged retarded/advanced Green function in the absence of inhomogeneity 
(i.e., averaged with the homogeneous part of $p_{\text{imp}} ({\bm r})$) is given by
\begin{align}
G^{\text R/ \text A}_{\bm k}(\epsilon ) 
= \frac{1}{\epsilon +\mu -\epsilon_{\bm k} \pm i\gamma },
\end{align}
 where $\gamma = \pi n_{\text i}u_{\text i}^2 N(\mu ) (1+\frac{2}{3}\lambda_{\text{so}}^2k_{\text F}^4)$ is the damping rate, 
with the Fermi-level density of states (per spin) $N(\mu ) = \frac{mk_{\text F}}{2\pi^2}$ 
 for 3D, $N(\mu ) = \frac{m}{2\pi}$ for 2D, 
the Fermi wave number $k_{\text F}$, and the chemical potential $\mu$.

 We apply a spatially-modulated, time-dependent electromagnetic field, 
$A_{\nu }(\bm r, t ) = A_{\nu }(\bm q, \omega )  \, e^{i({\bm q} \cdot {\bm r} - \omega t)}$, 
where $\bm q$ is the wave vector, $\omega$ is the frequency, 
and the four-vector notation $A_{\nu} = (-\phi, \bm A)$ has been used. 
 The perturbation Hamiltonian is given by 
\begin{align}
 H_{\text{ext}} = -\hat{j}_{\text e, \nu}(-\bm q) A_{\nu}(\bm q, \omega ) \, e^{-i\omega t} ,
\label{eq:Hext}
\end{align}
where $\hat{j}_{\text e, \nu}
=(\hat{\rho}_{\text e}, \hat{\bm j}_{\text e})
$ is the electric charge/current density operator. 
 Throughout this paper, we assume that 
the spatial variations of $n_{\text i}(\bm r)$ and $A_{\nu}$ are 
much slower compared to the electron mean free path $l$, 
and the temporal variation of $A_{\nu}$ is much slower compared to the electron damping rate $\gamma$. 
 These are expressed by $q, Q  \ll l^{-1}$ and $\omega \ll \gamma$.

 The electric charge- and electric current-density operators are given by
\begin{align}
\hat j_{\text e, 0}(\bm q) &= \hat \rho_{\text e}(\bm q) 
= -e \sum_{\bm k} \psi^{\dagger}_{\bm k-\frac{\bm q}{2}} \psi_{\bm k+\frac{\bm q}{2}},
\\
\hat j_{\text e, i} (\bm q) &= -e \sum_{\bm k} v_i \psi^{\dagger}_{\bm k-\frac{\bm q}{2}} \psi_{\bm k+\frac{\bm q}{2}}
+ \hat j^{\text a}_{\text e, i}(\bm q)
,
\end{align}
and the spin- and spin-current-density operators are given by
\begin{align}
\hat{j}^{\alpha}_{\text s, 0}(\bm q) &= \hat{\sigma }^{\alpha}(\bm q) = \sum_{\bm k} \psi^{\dagger}_{\bm k-\frac{\bm q}{2}} \sigma^{\alpha} \psi_{\bm k+\frac{\bm q}{2}},
\\
\hat{j}^{\alpha}_{\text s, i}(\bm q) &= \sum_{\bm k}v_i \psi^{\dagger}_{\bm k-\frac{\bm q}{2}} \sigma^{\alpha} \psi_{\bm k+\frac{\bm q}{2}}
+\hat{j}^{\text a, \alpha}_{\text s, i}(\bm q),
\end{align}
where $\alpha$ ($= x, y, z$) specifies the spin direction, 
and $i$ ($=x, y, z$) specifies the current direction. 
 The second terms ($\hat j^{\text a}_{\text e, i}$ and $\hat j^{\text a, \alpha}_{\text s, i}$) 
of the charge and spin currents are the ``anomalous" parts,  
\begin{align}
 \hat j^{\text a}_{\text e, i}(\bm q) 
&= - i e \lambda_{\text{so}} 
\sum_{\bm k, \bm k'} V_{\bm k-\bm k'} \psi^{\dagger}_{\bm k-\frac{\bm q}{2}} [\bm \sigma \times (\bm k-\bm k')]_i \psi_{\bm k'+\frac{\bm q}{2}}
,
\\
 \hat{j}^{\text a, \alpha}_{\text s, i} (\bm q) 
&= -i\lambda_{\text{so}} \epsilon_{\alpha ij}  
 \sum_{\bm k, \bm k'} V_{\bm k-\bm k'} (\bm k-\bm k')_j 
  \psi^{\dagger}_{\bm k-\frac{\bm q}{2}} \psi_{\bm k'+\frac{\bm q}{2}},
\end{align}
where $\epsilon_{\alpha ij}$ is the Levi-Civita symbol

\subsection{Inclusion of Coulomb interaction}
\label{model with Coulomb}

 Because of the inhomogeneity of impurities, electron density can be inhomogeneous. 
 In addition, the inhomogeneous external perturbation [Eq.~(\ref{eq:Hext})] 
can also cause charge accumulation. 
 To treat such situations properly, we consider the long-range Coulomb interaction, 
\begin{align}
 H_{\text C} = \frac{1}{2} \sum_{\bm K \neq 0} U_{\bm K} 
     \sum_{\bm k, \bm k'} \sum_{\sigma, \sigma'} 
     \psi^{\dagger}_{{\bm k-\bm K}, \sigma} \psi^{\dagger}_{{\bm k'+\bm K}, \sigma'} 
     \psi_{{\bm k'} \sigma'} \psi_{{\bm k} \sigma}, 
\label{eq:Coulomb}
\end{align}
where $U_{\bm K} = \frac{e^2}{\epsilon_0 K^2}$ for 3D ($U_{\bm K} = \frac{e^2}{2\epsilon_0 K}$ for 2D) 
with $\epsilon_0$ being the dielectric constant in the medium considered. 
 The analysis will be done in Sec.~\ref{with} based on the total Hamiltonian, 
$H + H_{\text C}$, where $H$ is given by Eq.~(\ref{H}).

\section{calculation}
\label{calculation}

 We are interested in the spin accumulation and spin current  
that arise in linear response\cite{kubo} to $A_\nu$ (with wave vector ${\bm q}$), 
and at the first order in $\delta n_{\rm i}$ (with wave vector ${\bm Q}$), 
\begin{align}
 \langle \hat{j}^{\alpha}_{\text s, \mu}(\bm q+\bm Q)\rangle_\delta 
= [ K^{\text{sj}, \alpha}_{\mu \nu} + K^{\text{ss}, \alpha}_{\mu \nu}]  A_{\nu}(\bm q, \omega ) . 
\label{eq:js_Q+q}
\end{align}
 The suffix $\delta$ indicates that this is first order in $\delta n_{\rm i}$. 
 We will also present the terms zeroth-order in $\delta n_{\rm i}$, 
which will be denoted by $\langle \cdots \rangle_0$. 
 As indicated in Eq.~(\ref{eq:js_Q+q}), the response function arises from two processes\cite{karplus1, smit1, smit2, berger1, bruno1, wang1, wang2, raimondi1}. 
 One, denoted by $K^{\text{sj}, \alpha}_{\mu \nu}$, comes from the side-jump type processes, 
and the other, denoted by $K^{\text{ss}, \alpha}_{\mu \nu}$, results from the skew-scattering 
type processes. 
 We note that equilibrium spin currents do not arise.

\subsection{Charge channel}

 Before proceeding, we first look at the charge and charge-current densities. 
 Those in a homogeneous system (terms zeroth order in $\delta n_{\rm i}$ 
but first order in $A_{\nu }(\bm q, \omega )$) are given by 
\begin{align}
 \langle \hat{\rho}_{\text e} \rangle_0 &= -\sigma_{\text c} \frac{i\bm q\cdot \bm E}{Dq^2-i\omega } , 
\label{cd_0}
\\
 \langle \hat{j}_{\text e, i} \rangle_0 
&= \sigma_{\text c} \left[ \delta_{ij} - \frac{D q_i q_j}{Dq^2-i\omega } \right] E_j , 
\label{cc_0}
\end{align}
where ${\bm E} = - \nabla \phi - \partial_t {\bm A}$ is the applied electric field,  
$\sigma_{\text c} = n_{\text e}e^2 \tau /m$ is the 
Drude conductivity 
and $D=v_{\text F}^2 \tau /3$ is the diffusion constant   
with the electron number density $n_{\text e}=\frac{2m}{3}v_{\text F}^2N(\mu )$, 
the Fermi velocity $v_{\text F}$, and the relaxation time $\tau =(2\gamma )^{-1}$.
 The terms first order in $\delta n_{\rm i}$ are calculated as
\begin{align}
 \average{\hat \rho_{\text e}}_{\delta} 
&= \sigma_{\text c} \frac{\delta n_{\text i}(\bm Q)}{n_{\text i}} 
     \frac{1}{D(\bm q+\bm Q)^2-i\omega} 
\nonumber \\ & \ \ 
\times \left[ 
i(q_j+Q_j) 
- \frac{D\bm q\cdot (\bm q+\bm Q)iq_j}{Dq^2-i\omega} 
\right] E_j,
\label{cd}
\\
 \average{\hat j_{\text e, i}}_{\delta} 
&= -\frac{\delta n_{\text i}(\bm Q)}{n_{\text i}} \sigma_{\text c}E_i -\delta D (\bm Q) iq_i  \average{\hat \rho_{\text e}}_{0} 
\nonumber \\ & \ \ 
\hspace{30mm}
-Di(q_i+Q_i) \average{\hat \rho_{\text e}}_{\delta} . 
\label{cc}
\end{align}
where 
$\delta D(\bm Q)=-\frac{\delta n_{\text i}(\bm Q)}{n_{\text i}}D$ 
is the modification of the diffusion coefficient. 
 We note that these satisfy the continuity equation, 
$\partial_t \average{\hat \rho_{\text e}}_{\delta} + \nabla \cdot \average{\hat{\bm j}_{\text e}}_{\delta}=0$.

\subsection{Side-jump process}

\begin{figure}[t]
 \begin{center}
  \includegraphics[clip,width=85mm]{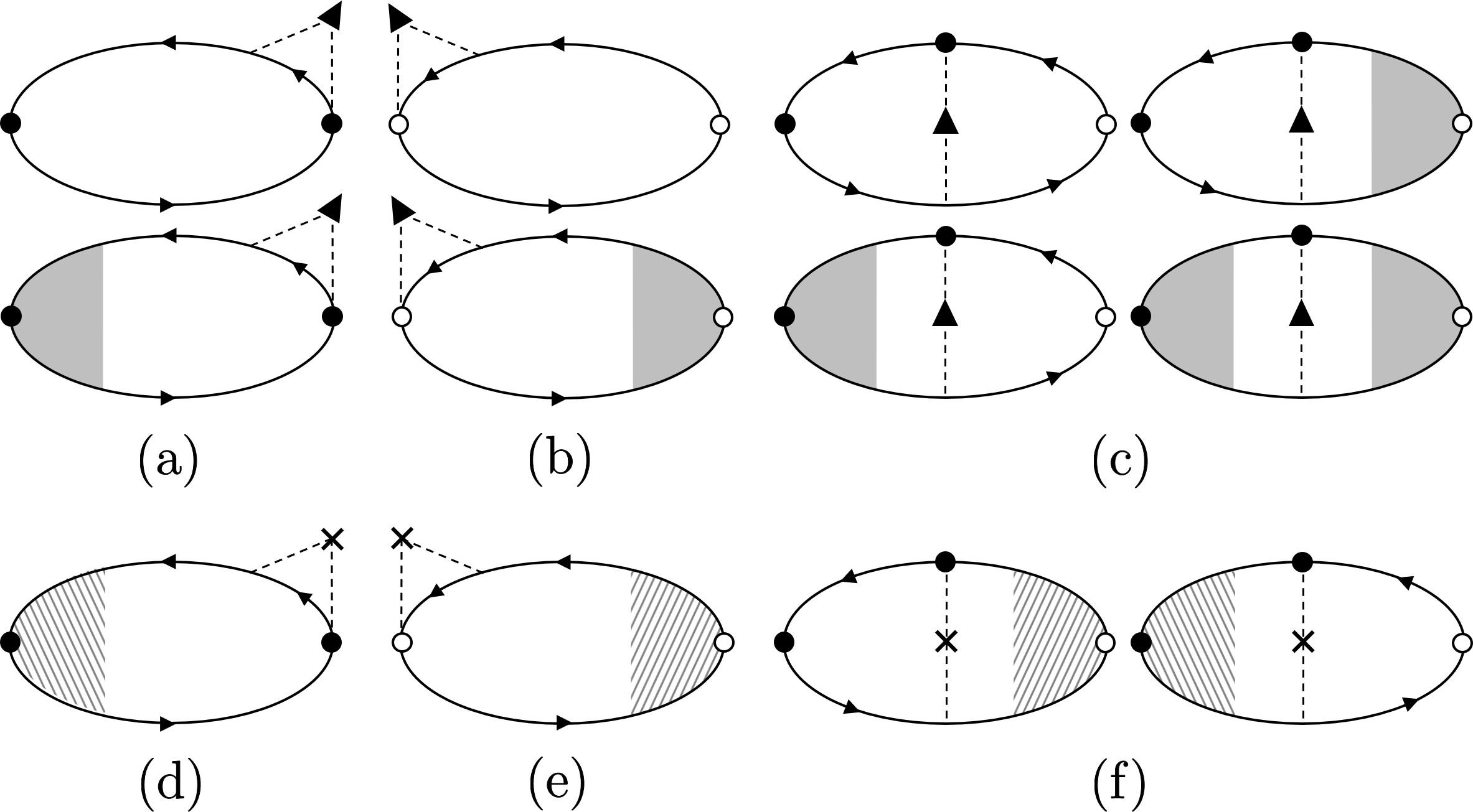}
  \caption{Side-jump type contribution to the spin current and spin accumulation. 
 In each diagram, the left vertex represents spin current or spin density, 
and the right vertex represents electric current. 
 The black circles represent spin-flip or spin-dependent vertices accompanied by Pauli matrices, 
and the white circles represent spin-independent vertices. 
 The cross with a dashed line represents an impurity with homogeneous distribution, 
and the triangle with a dashed line represents the inhomogeneous part of the impurity distribution. 
 The contributions of (a) to (c) are due to the inhomogeneous  SOI scattering 
and those of (d) to (f) are due to the inhomogeneous  normal scattering. 
 The shaded and hatched regions represent impurity ladder vertex corrections (see Appendix A), 
the former (latter) being zeroth (first) order in $\delta n_{\rm i}$. 
 The latter is shown in Fig.~\ref{vc_tri}.
 The diagrams (a) and (d)  come from the ``anomalous" part of electric current, 
and those of (b) and (e) come from the ``anomalous" part of spin current.
 The upside-down diagrams are also considered in the calculation.}
  \label{sj_mod}
 \end{center}
\end{figure}

 We now focus on the spin channel, and calculate spin accumulation and spin-current density. 
 The side-jump type contributions are characterized by a $\tau$-independent SH conductivity, 
or the SH angle,  
$\alpha^{\text{sj}}_{\text{SH}} = - 2m\lambda_{\text{so}} /\tau$, inversely proportional to $\tau$. 
 The terms first order in $\delta n_{\rm i}$ are classified into two types, 
one due to the inhomogeneity of  SOI scattering, 
and the other due to the inhomogeneity of  normal scattering.
 The former is expressed by the diagrams shown in Fig.~\ref{sj_mod}~(a), (b) and (c), 
and each set of diagram gives 
\begin{align}
 K^{\text{sj(a)},\alpha}_{\mu j} 
&= -e\lambda_{\text{so}} \delta n_{\text i}(\bm Q) u_{\text i}^2 \frac{\omega}{\pi} \epsilon_{\alpha jl} 
\nonumber \\
&\times
 \sum_{\bm k, \bm k'} v_{\mu} k_l  G^{\text R}_{\bm k ++} G^{\text A}_{\bm k --}
\left(
G^{\text R}_{\bm k' +} - G^{\text A}_{\bm k' -}
\right) ,
\\
K^{\text{sj(b)},\alpha}_{ij} &= -e\lambda_{\text{so}} \delta n_{\text i}(\bm Q)u_{\text i}^2 \frac{\omega}{\pi} \epsilon_{\alpha il} 
\nonumber
\\
&\times
 \sum_{\bm k, \bm k'} v_j k_lG^{\text R}_{\bm k + } G^{\text A}_{\bm k - }
\left(
G^{\text R}_{\bm k' + } - G^{\text A}_{\bm k' - }
\right) ,
\\
K^{\text{sj(c)},\alpha}_{\mu j} &= -e\lambda_{\text{so}} \delta n_{\text i}(\bm Q)u_{\text i}^2 \frac{\omega}{\pi} \sum_{\bm k, \bm k'} v_{\mu} v_j  [ \bm k' \times (\bm q+\bm Q) 
\nonumber \\
&
- \bm k\times \bm q  ]_{\alpha}
G^{\text R}_{\bm k + +} G^{\text A}_{\bm k - -}
G^{\text R}_{\bm k' + } G^{\text A}_{\bm k' - }, 
\end{align}
where
 $G^{\text{R/A}}_{\bm k ss'}=G^{\text{R/A}}_{\bm k + s\frac{\bm q}{2}+s'\frac{\bm Q}{2}}(s\frac{\omega}{2})$ 
 and $G^{\text{R/A}}_{\bm k s}=G^{\text{R/A}}_{\bm k+s\frac{\bm q}{2}}(s \frac{\omega}{2})$ 
with $s, s'=\pm $. 
 The contributions from the inhomogeneous normal  scattering  are given  
in Fig.~\ref{sj_mod}(d)-(f), and further details are described in Appendix~\ref{app_side-jump}.

\begin{figure}[t]
 \begin{center}
  \includegraphics[clip,width=70mm]{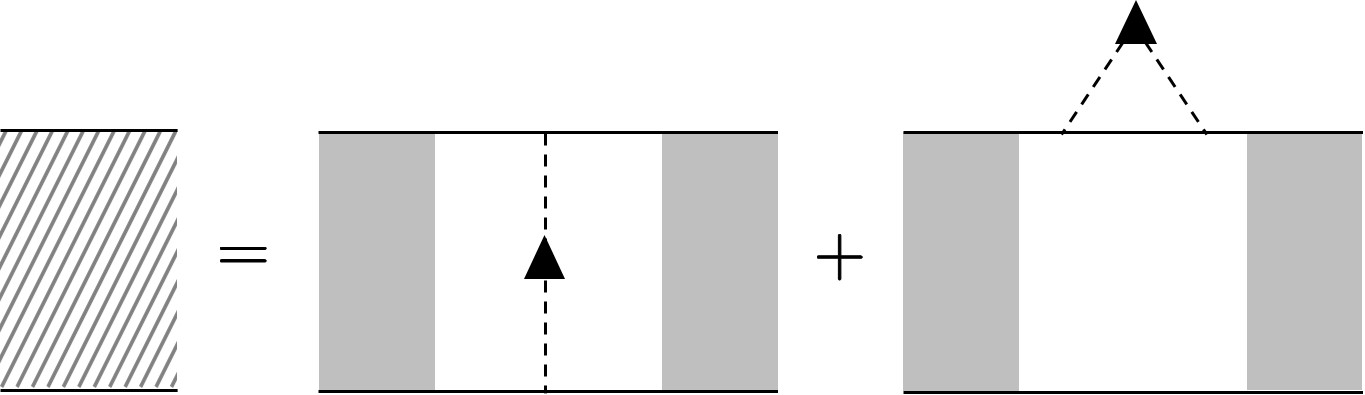}
  \caption{The hatched rectangle (left-hand side) represents the inhomogeneous part 
(first order in $\delta n_{\rm i}$) of the four-point impurity-ladder vertex. 
 Only two typical diagrams are shown on the right-hand side, 
but diagrams without the shaded rectangle (the homogeneous part of impurity-ladder diagrams, 
see Fig.~\ref{vcs}) 
on either or both sides should also be included.}
  \label{vc_tri}
 \end{center}
\end{figure}

 With these diagrams, and also by including ladder vertex corrections, 
the spin accumulation and spin-current density have been obtained as
\begin{align}
  \average{\hat \sigma^{\alpha}}_{\delta }^{\text{sj}} 
=&  \frac{\alpha^{\text{sj}}_{\text{SH}}}{-e}  
\frac{\delta D (\bm Q) \, \bm q \!\cdot\! (\bm q+\bm Q) }{D(\bm q+\bm Q)^2-i\omega +\tau_{\text{sf}}^{-1}} 
\nonumber \\ & 
\hspace{25mm}
\times \frac{[i\bm q\times \average{\hat{\bm j}_{\text e}}_0]_{\alpha}}{Dq^2-i\omega + \tau_{\text{sf}}^{-1}},
\label{spin_sj}
\\
  \average{\hat j^{\alpha}_{\text s, i}}_{\delta}^{\text{sj}} =&
-\alpha^{\text{sj}}_{\text{SH}}\epsilon_{\alpha ij} Di(q_j+Q_j) \frac{\average{\hat \rho_{\text e}}_{\delta}}{-e} 
\nonumber
\\
&
 - Di(q_i+Q_i) \average{\hat \sigma^{\alpha}}_{\delta  }^{\text{sj}} 
- \delta D(\bm Q) iq_i  \average{\hat \sigma^{\alpha}}_{0}^{\text{sj}} 
,
 \label{sc_sj}
\end{align}
where 
$\tau_{\text{sf}} = \frac{9}{8}(\lambda_{\text{so}}^2k_{\text F}^4)^{-1}\tau$ is the spin relaxation time. 
 These are to be added to those in the absence of inhomogeneity $\delta n_{\text i}$,
 \cite{wang1, wang2, raimondi1, tatara1} 
\begin{align}
 \average{\hat \sigma^{\alpha}}_{0}^{\text{sj}} 
&= -\alpha^{\text{sj}}_{\text{SH}} \frac{\sigma_{\text c}}{-e} 
   \frac{ [i\bm q\times \bm E]_\alpha}{Dq^2-i\omega + \tau_{\text{sf}}^{-1}} , 
 \label{spin_sj_0}
\\
 \average{\hat j^{\alpha}_{\text s, i}}_{0}^{\text{sj}} 
&= \alpha^{\text{sj}}_{\text{SH}}\epsilon_{\alpha ij} 
     \average{\hat{j}_{\text e, j}}_0/(-e) -Diq_i \average{\hat \sigma^{\alpha}}_{0}^{\text{sj}} . 
 \label{sc_sj_0}
\end{align}

The spin accumulation $\average{\hat \sigma^{\alpha}}_{\delta }^{\text{sj}}$ [Eq.~(\ref{spin_sj})] 
is due to the modulation of diffusion coefficient $D$ $(\propto n_{\text i}^{-1})$ 
in the denominator of $\average{\hat \sigma^{\alpha}}_{0}^{\text{sj}}$ [Eq.~(\ref{spin_sj_0})].
 Note that they vanish when the electric field is uniform ($\bm q=0$).
 In the spin current $\average{\hat j^{\alpha}_{\text s, i}}_{\delta}^{\text{sj}}$ [Eq.~(\ref{sc_sj})], 
the first term is the spin Hall current arising from the diffusion charge current 
[the last term in Eq.~(\ref{cc})], 
and the second and third terms are the diffusion spin current whose dependence on $\delta n_{\text i}$ 
is through the modulation of spin accumulation $\average{\hat \sigma^{\alpha}}_{\delta}^{\text{sj}}$ 
[Eq.~(\ref{spin_sj})] and the modulation of diffusion coefficient $\delta D$, respectively.
 On the other hand, we found no spin Hall current originating from the first two terms of 
the charge current $\average{\hat j_{\text e, i}}_{\delta}$ [Eq.~(\ref{cc})].
These results are consistent with the fact that the SH conductivity due to side-jump process is independent of the impurity concentration, $\alpha^{\text{sj}}_{\text{SH}}\sigma_{\text c} \propto n_{\text i}^0$.
 Finally, the \lq\lq inhomogeneous contributions'', 
$\average{\hat \sigma^{\alpha}}_{\delta }^{\text{sj}} $ 
and $\average{\hat j^{\alpha}_{\text s, i}}_{\delta}^{\text{sj}}$ [Eqs.~(\ref{spin_sj}) and (\ref{sc_sj})] 
satisfy the same continuity equation (with spin-relaxation term), 
\begin{align}
   \partial_t \average{\hat \sigma^{\alpha}}_{\delta}^{\text{sj}} 
 + \nabla \cdot \average{\hat{\bm j}^{\alpha}_{\text s}}_{\delta}^{\text{sj}} 
= -\frac{\average{\hat \sigma^{\alpha}}_{\delta }^{\text{sj}}}{\tau_{\text{sf}}},
\label{eq:cont_delta_sj}
\end{align}
as that of the \lq\lq homogeneous contributions''.

\subsection{Skew scattering process} 

\begin{figure}[t]
 \begin{center}
  \includegraphics[clip,width=80mm]{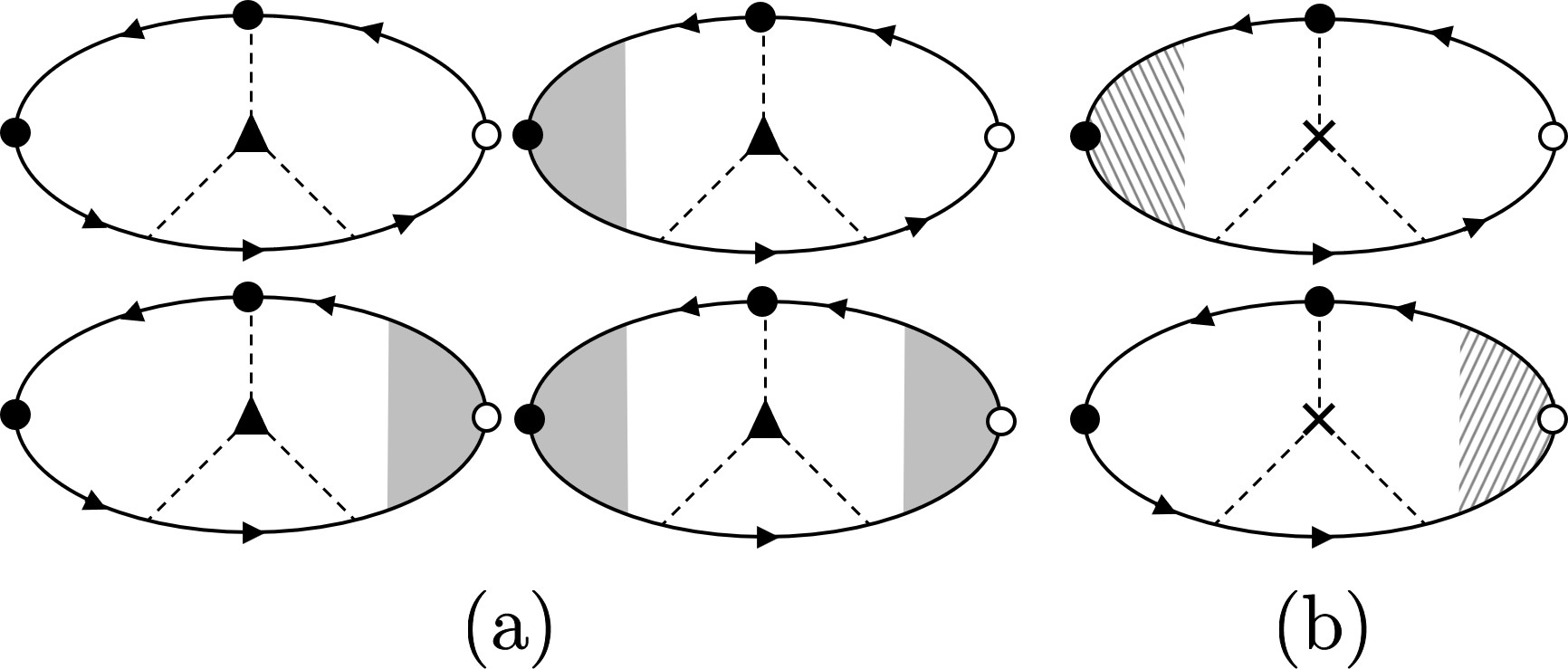}
  \caption{Skew-scattering type contribution to the spin current and spin accumulation 
in the first order in the inhomogeneity $\delta n_{\rm i}$.  
 The inhomogeneity comes either from the  SOI scattering (a) 
and from the normal  scattering  (b).
The upside-down diagrams are also considered in the calculation.
 }
  \label{skew_mod}
 \end{center}
\end{figure}

The skew-scattering type contributions are characterized by the diagrams 
as shown in Fig.~\ref{skew_mod}. 
 Again, the terms first order in $\delta n_{\rm i}$ are classified into two types  
according to whether the inhomogeneity comes from the SOI  scattering  
or normal  scattering . 
 The response functions of the former type (SOI inhomogeneity) 
are expressed by the diagrams in Fig~\ref{skew_mod}(a), the first of which leads to  
\begin{align}
K^{\text{ss}, \alpha}_{\mu j} &= e\lambda_{\text{so}} \delta n_{\text i}(\bm Q) u_{\text i}^3 \frac{\omega}{\pi} \sum_{\bm k, \bm k', \bm k''} v_{\mu} v_j' 
[\bm k\times \bm k']_{\alpha} 
\nonumber
\\ & \times
\left[ G^{\text R}_{\bm k''}\left( \frac{\omega}{2}\right) - G^{\text A}_{\bm k''}\left( -\frac{\omega}{2} \right) \right]  
G^{\text R}_{\bm k++} G^{\text A}_{\bm k--}
G^{\text R}_{\bm k+} G^{\text A}_{\bm k-}.
\end{align}
The latter type (normal-scattering inhomogeneity) is described by the diagrams 
in Fig~\ref{skew_mod} (b). 
 More details can be found in Appendix~\ref{app_skew_scattering}.

 These diagrams, with ladder vertex corrections included, lead to the following expression of 
the spin accumulation and spin-current density, 
\begin{align}
 \average{\hat \sigma^{\alpha}}_{\delta }^{\text{ss}} 
=& - \frac{\alpha^{\text{ss}}_{\text{SH}}}{-e} 
\frac{1}{D(\bm q+\bm Q)^2-i\omega + \tau_{\text{sf}}^{-1}} 
\nonumber
\\ &
\times 
\biggl\{ 
\left[ i(\bm q+\bm Q) \times \average{\hat{ \bm j}_{\text e}}_{\delta} \right]_{\alpha} 
\nonumber
\\ & 
\ \ \ \ \  
- \frac{\delta D \, \bm q \!\cdot\! (\bm q+\bm Q) [i\bm q\times \average{\hat{\bm j}_{\text e}}_0]_{\alpha}}{Dq^2-i\omega + \tau_{\text{sf}}^{-1} }
\biggr\}
,
\label{spin_ss}
\\
\average{\hat j^{\alpha}_{\text s, i}}_{\delta }^{\text{ss}} 
=& \alpha^{\text{ss}}_{\text{SH}} \epsilon_{\alpha ij} \frac{\average{\hat j_{\text e, j}}_{\delta} }{-e} 
- Di(q_i+Q_i) \average{\hat \sigma^{\alpha}}_{\delta  }^{\text{ss}}
\nonumber \\ &
-\delta D(\bm Q)iq_i  \average{\hat \sigma^{\alpha}}_{0}^{\text{ss}},
\label{sc_ss}
\end{align}
where $\alpha^{\text{ss}}_{\text{SH}} 
 = \frac{2\pi}{3} k_{\text F}^2 \lambda_{\text{so}} N(\mu )u_{\text i}$ 
is the SH angle due to skew scattering. 
 These results are to be added to the homogeneous contributions,\cite{tatara1}  
\begin{align}
 \average{\hat \sigma^{\alpha}}_{0}^{\text{ss}} 
&= -\alpha^{\text{ss}}_{\text{SH}}  \frac{\sigma_{\text c}}{-e}  
   \frac{[i\bm q\times \bm E]_{\alpha}}{Dq^2-i\omega + \tau_{\text{sf}}^{-1}} , 
\\
 \average{\hat j^{\alpha}_{\text s, i}}^{\text{ss}}_0 
&= \alpha^{\text{ss}}_{\text{SH}}\epsilon_{\alpha ij} 
    \average{\hat{j}_{\text e, j}}_0/(-e) -Diq_i \average{\hat \sigma^{\alpha}}_{0}^{\text{ss}} . 
\end{align}
which 
have exactly the same form as $\average{\hat \sigma^{\alpha}}_{0}^{\text{sj}}$ 
and $\average{\hat j^{\alpha}_{\text s}}^{\text{sj}}_0$ [Eqs.~(\ref{spin_sj_0}) and (\ref{sc_sj_0})] 
except for the coefficient ($\alpha_{\text{SH}}^{\text{ss}}$ instead of $\alpha_{\text{SH}}^{\text{sj}}$).

 The contributions (\ref{spin_ss}) and (\ref{sc_ss}) are written with the charge current 
$\average{\hat j_{\text e, i}}_{\delta}$ [Eq.~(\ref{cc})].
 Unlike the side-jump contribution, the spin accumulation (\ref{spin_ss}) remains finite at $\bm q = \bm 0$. 
 Thus a uniform electric field ${\bm E}$ induces a spin accumulation through the inhomogeneous skew scattering. 
In Eq.~(\ref{sc_ss}), the first term is the spin Hall current arising from the charge current 
$\average{\hat j_{\text e, i}}_{\delta}$ [Eq.~(\ref{cc})], 
while the remaining terms are diffusion spin current similar to those in the side-jump contribution 
$\average{\hat j^{\alpha}_{\text s, i}}_{\delta}^{\text{sj}}$ [Eq.~(\ref{sc_sj})].
 These results are consistent with the fact that the SH angle due to skew scattering  
is independent of the impurity concentration, thus 
$\alpha^{\text{ss}}_{\text{SH}}\sigma_{\text c}\propto n_{\text i}^{-1}$.
 Finally, the modulated parts satisfy the same (continuity) equation as Eq.~(\ref{eq:cont_delta_sj}), 
\begin{align}
\partial_t \average{\hat \sigma^{\alpha}}_{\delta }^{\text{ss}} + \nabla \cdot \average{\hat{\bm j}^{\alpha}_{\text s}}_{\delta }^{\text{ss}} 
= -\frac{\average{\hat \sigma^{\alpha}}_{\delta }^{\text{ss}}}{\tau_{\text{sf}}}  . 
\end{align}
 Therefore, the total spin accummlation, 
$\average{\hat \sigma^{\alpha}}_{\delta }
 = \average{\hat \sigma^{\alpha}}_{0 }^{\text{sj}} 
  + \average{\hat \sigma^{\alpha}}_{0 }^{\text{ss}} 
  + \average{\hat \sigma^{\alpha}}_{\delta }^{\text{sj}} 
  + \average{\hat \sigma^{\alpha}}_{\delta }^{\text{ss}} $ 
and the total spin current (defined similarly) also satisfy the same equation.

\subsection{Extrinsic Rashba process \label{ex-rashba}}

\begin{figure}[t]
 \begin{center}
  \includegraphics[width=85mm]{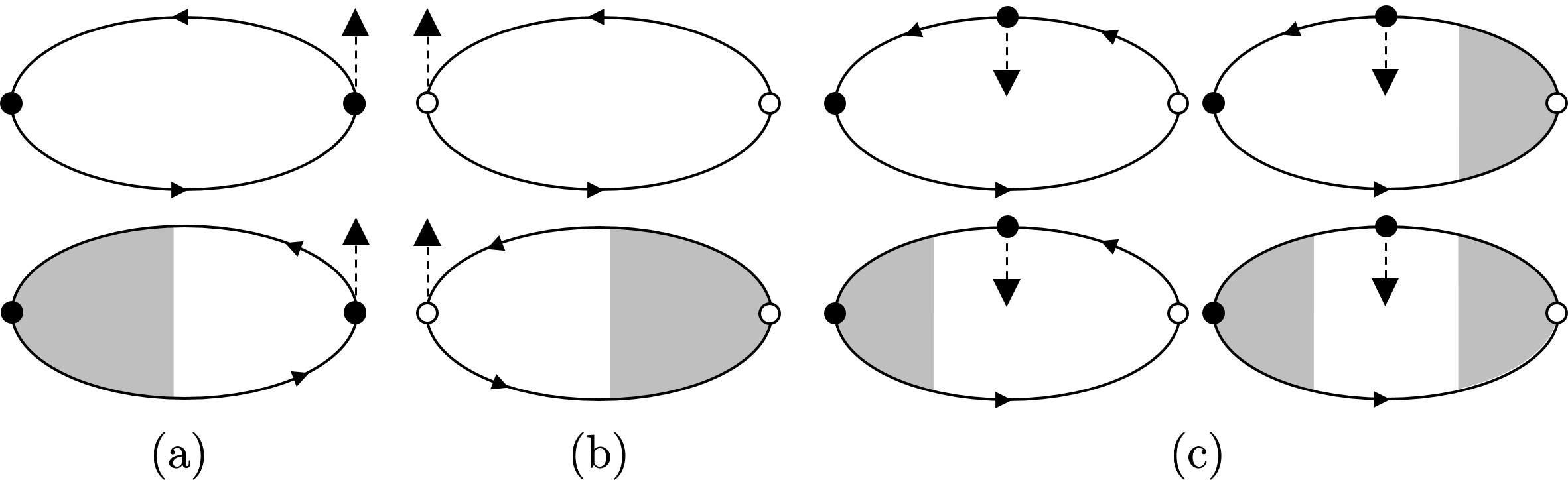}
  \caption{
 Rashba type contribution to the spin current and spin accumulation which are first order 
in the impurity potential strength $u_{\text i}$.
The diagrams (a) and (b) come from the anomalous part of the electric current and spin current operators, 
respectively. 
 The diagrams without vertex corrections (i.e., without shaded regions) in (a), (b), and (c) are canceled out. 
 The diagrams with left vertex corrections in (a) and (c) are canceled out, 
and so are the diagrams with right vertex corrections in (b) and (c). 
 The diagram in (c) having vertex corrections on both sides vanishes 
up to second order in $\bm q$ and $\bm Q$. } 
  \label{1st}
 \end{center}
\end{figure}

 Because of the modulated distribution of SOI impurities, 
the SOI Hamiltonian survives the impurity average, 
\begin{align}
\average{H_{\text{so}}}_{\text{av}} 
= \lambda_{\text{so}}\delta n_{\text i}(\bm Q)u_{\text i} \sum_{\bm k} (i\bm Q \times \bm k) \cdot 
   \psi^{\dagger}_{\bm k+\bm Q} \bm \sigma \psi_{\bm k} . 
\label{eq:H_ex_Rashba}
\end{align}
 This may be called ``extrinsic Rashba SOI" since it arises from the inversion symmetry breaking 
by the impurity distribution.
 However,  the contribution to the spin accumulation and spin current, 
shown in Fig.~\ref{1st}, turned out to vanish. 
 This is in agreement with the well-known fact that the Edelstein effect\cite{edelstein} 
vanishes in such perturbative calculation.
 To obtain the spin accumulation due to the Edelstein effect via the extrinsic Rashba SOI, 
we need to treat it nonperturbatively. 
 If we neglect the momentum change 
($\psi^{\dagger}_{\bm k+\bm Q} \bm \sigma \psi_{\bm k} \to \psi^{\dagger}_{\bm k} \bm \sigma \psi_{\bm k}$), 
the spin accumulation is obtained as
\begin{align}
  \average{\hat \sigma^{\alpha}}^{\text{R}}  
= - \lambda_{\text{so}} u_{\text i} \frac{3\sigma_{\text c}}{ev_{\text F}^2} 
    [ \nabla \delta n_{\text i}(\bm r) \times \bm E]_{\alpha}  , 
\end{align}
in 3D. 
 This spin accumulation will in principle contribute to the charge-to-spin conversion. 
 However, its magnitude relative to Eq.~(\ref{spin_ss}) 
(with ${\bm q} = {\bm 0}$ and $\omega=0$), estimated as 
\begin{align}
  \frac{\average{\hat \sigma^{\alpha}}^{\rm R} }{\average{\hat \sigma^{\alpha}}_{\delta }^{\rm ss} } 
&\simeq   \frac{(DQ^2 + \tau_{\text{sf}}^{-1}) \tau}{\epsilon_{\rm F} \tau}  \cdot  
     \frac{n_{\rm i}}{n_{\rm e}} , 
\end{align}
is much smaller than unity. 
(Note that $DQ^2 \tau$, $\tau/\tau_{\text{sf}}$, $n_{\rm i}/n_{\rm e}$ and 
$(\epsilon_{\rm F} \tau)^{-1}$ are all small quantities.) 
 Therefore, this contribution can safely be neglected.

 It was reported that in a 2D magnetic Rashba system diagrams with crossed impurity lines contribute 
by the same order to the anomalous Hall effect\cite{ado}. 
 In our model, however, even if we assume 2D, 
such contributions are higher order in the damping $\gamma$.

\subsection{Zeeman process}

 When the electromagnetic field has finite ${\bm q}$, there is a magnetic field which also couples to spin 
(Zeeman coupling).

 For a homogeneous system, $\bm Q = \bm 0$, 
the spin accumulation induced by the Zeeman coupling is calculated as \cite{funato}
\begin{align}
 \average{\sigma^{\alpha}}^{\text{Z}}_0  
 = -\frac{e N(\mu )}{m} 
    \left[ 1+\frac{i\omega}{Dq^2-i\omega +\tau^{-1}_{\text{sf}}} \right] B^{\alpha} . 
\label{s_Zeeman}
\end{align}
 The first term is simply the Pauli paramagnetic response. 
 According to Maxwell equation, the second term is proportional to $\nabla \times \bm E$, 
and has the same form as 
$\average{\sigma^{\alpha}}_{\text{SH}} 
  = \average{\sigma^{\alpha}}^{\text{sj}}_0+\average{\sigma^{\alpha}}^{\text{ss}}_0$. 
 The ratio of the spin accumulation induced by the spin Hall effect 
and the one caused by the Zeeman coupling (second term) is 
\begin{align}
 \frac{\average{\sigma^{\alpha}}_{\text{SH}}}{\average{\sigma^{\alpha}}^{\text{Z, 2nd}}_0 } 
   = \frac{2}{3} \frac{\epsilon_{\text F} \tau}{\hbar} \alpha_{\text{SH}} . 
\label{s_Zeeman_ratio}
\end{align}
 Since $\alpha_{\text{SH}} \ll 1$ and $\epsilon_{\text F} \tau/\hbar \gg 1$, 
this ratio can be greater or less than unity. 
 In very good metals with large SOI, namely, 
for $\alpha_{\text{SH}} (\epsilon_{\text F} \tau/\hbar) \gg 1$, 
the Zeeman contribution can be neglected, but it can not in the opposite case. 
 This holds also for inhomogeneous systems with $\bm Q \neq 0$. 
 In this paper, we focus on the case, $\alpha_{\text{SH}} (\epsilon_{\text F} \tau/\hbar) \gg 1$, 
in which the Zeeman coupling can be neglected.

\section{Results and application}

 In this section, we summarize the results obtained in the preceding section (without Coulomb interaction), 
and discuss the equation that governs them. 
 For illustration, we apply the equation to a thin film.  
 The effects of long-range Coulomb interaction will be studied in the next section (Sec.~\ref{with}).

\subsection{Summary of results}

 In the present model, there are two origins of spatial modulation of the currents, 
impurity distribution (with wave vector ${\bm Q}$) and the applied electric field (with wave vector ${\bm q}$).

 Let us first consider the case that a uniform electric field ($\bm q = \bm 0$) is applied 
to the inhomogeneous system ($\bm Q \ne \bm 0$).
 The spin accumulation and spin current are given by
\begin{align}
 \average{\hat \sigma^{\alpha}(\bm Q)}_{\delta} 
&= -\frac{\alpha^{\text{ss}}_{\text{SH}}}{-e} \frac{[i\bm Q \times \average{\bm j_{\text e}(\bm Q)}_{\delta}]_{\alpha}}{DQ^2 - i\omega + \tau_{\text{sf}}^{-1}}, 
\label{eq:s^y_Q}
\\
  \average{\hat j^{\alpha}_{\text s, i}(\bm Q)}_{\delta} 
&= \alpha^{\text{ss}}_{\text{SH}} \epsilon_{\alpha ij} 
    \frac{\average{  j_{{\text e},j}(\bm Q)}_{\delta}}{-e} - DiQ_i 
    \average{\sigma^{\alpha} (\bm Q)}_{\delta}
\nonumber
\\
& -\alpha^{\text{sj}}_{\text{SH}} \epsilon_{\alpha ij} DiQ_j 
    \frac{\average{\rho_{\text e}(\bm Q)}_{\delta}}{-e} , 
\label{eq:js^y_Q}
\end{align}
where $\average{\rho_{\text e}(\bm Q)}_{\delta}$ and $\average{\bm j_{\text e}(\bm Q)}_{\delta}$ are given, 
respectively, by Eqs.~(\ref{cd}) and (\ref{cc}) with $\bm q=0$.

They satisfy the spin diffusion equation, 
\begin{align}
( D\nabla^2 -\partial_t - \tau_{\text{sf}}^{-1}) \, \average{ \sigma^{\alpha} } 
= \alpha^{\text{ss}}_{\text{SH}}  \Omega_{\text e}^{\alpha} ,
\label{diff1}
\end{align}
where $\bm \Omega_{\text e} = \frac{1}{-e} \nabla \times \bm j_{\text e}$ is the vorticity  
of the electron flow.
The right-hand side shows that the vorticity acts as a spin source via SHE. 
This may be considered as the ``effective SVC" in laboratory (inertial) frame, 
and this is caused by the spin-orbit coupling, or more specifically, by the skew-scattering process.
The side-jump contribution is absent because it is independent of impurity concentration, 
namely, $\average{\hat j^\alpha_{{\rm s}, i}}^{\rm sj}$ is spatially uniform (if ${\bm E}$ is uniform) 
even if $n_{\rm i}({\bm r})$ is inhomogeneous.

 When a spatially modulated electric field ($\bm q \ne \bm 0$) is applied 
to a uniform system ($\bm Q = \bm 0$), 
the spin diffusion equation is given by\cite{tatara1}
\begin{align}
( D\nabla^2 - \partial_t - \tau_{\text{sf}}^{-1}) \, \average{ \sigma^{\alpha} } 
= \alpha_{\text{SH}} \frac{\sigma_{\text c}}{-e} [\nabla \times \bm E]_{\alpha},
\end{align}
with $\alpha_{\text{SH}} = \alpha_{\text{SH}}^{\text{sj}}+\alpha_{\text{SH}}^{\text{ss}}$.
The right-hand side is the spin source term, which can also be written as 
$\alpha_{\text{SH}} \Omega_{\text e}^{\alpha} $ 
in terms of the vorticity $\bm \Omega_{\text e} = \frac{1}{-e} \nabla \times \bm j_{\text e}$ 
of the electric current ${\bm j}_{\rm e} = \sigma_{\rm c} {\bm E}$ generated now 
by the nonuniform electric field. 
 Therefore, we can see a strong connection between the vorticity of the electron flow and the spin accumulation independently of their origin.

 For general cases with $\bm q\neq 0$ and $\bm Q\neq 0$, the sum of the inhomogeneous contribution, 
Eqs.~(\ref{spin_sj}), (\ref{sc_sj}), (\ref{spin_ss}), and (\ref{sc_ss}), 
and the homogeneous contribution, 
$\average{\hat \sigma^{\alpha}}_0^{\rm sj \, (ss)}$ and 
$\average{\hat j^{\alpha}_{\text s, i}}^{\rm sj \, (ss)}_0$,
leads to

\begin{widetext}
\begin{align}
\average{\hat \sigma^{\alpha}} =&
-  \Pi_{\text s}(\bm q) \frac{\alpha_{\text{SH}}}{-e} [ i\bm q \times  \average{\hat{\bm j}_{\text e}}_0 ]_{\alpha}
-\Pi_{\text s}(\bm q+\bm Q) \biggl\{ \frac{\alpha^{\text{ss}}_{\text{SH}}}{-e} [ i(\bm q +\bm Q) \times  \average{\hat{\bm j}_{\text e}}_{\delta} ]_{\alpha}
- \frac{\alpha_{\text{SH}}}{-e}  \delta D(\bm Q) \, 
   \bm q \cdot (\bm q+\bm Q) 
  \Pi_{\text s}(\bm q) [ i\bm q \times \average{\hat{\bm j}_{\text e}}_0 ]_{\alpha} \biggr\} ,
\label{spin_tot}
\\
\average{\hat j^{\alpha}_{\text s,i}} =& 
 \alpha_{\text{SH}} \epsilon_{\alpha ij}  \frac{\average{j_{\text e, j}}_0}{-e} 
+ \epsilon_{\alpha ij} \left[ \frac{\alpha_{\text{SH}}^{\text{ss}}}{-e} \average{j_{\text e, j}}_{\delta}
-\frac{\alpha^{\text{sj}}_{\text{SH}}}{-e}  Di(q_j+Q_j) \average{\hat \rho_{\text e}}_{\delta} \right]
- \left[ D+\delta D (\bm Q) \right] iq_i \average{\hat \sigma^{\alpha}}_0
-Di(q_i+Q_i) \average{\hat \sigma^{\alpha}}_{\delta} ,
\label{sc_tot}
\end{align}
\end{widetext}
where $\Pi_{\text s}(\bm q) = (Dq^2 - i\omega + \tau_{\text{sf}}^{-1})^{-1}$ is the spin diffusion propagator. 
 $\average{\hat{\bm j}_{\text e}}_0$, $\average{\hat{\bm j}_{\text e}}_{\delta}$, and 
 $\average{\hat \rho_{\text e}}_{\delta}$ are given by Eqs.~(\ref{cc_0}), (\ref{cc}), and (\ref{cd}), 
respectively. 
 $\average{\hat \sigma^{\alpha}}_0$ and $\average{\hat \sigma^{\alpha}}_{\delta}$ are given 
by the first term and the remaining terms, respectively, in Eq.~(\ref{spin_tot}). 
 The first terms in Eqs.~(\ref{spin_tot}) and (\ref{sc_tot}) are zeroth order in $\delta n_{\text i}$, 
and the remaining terms are first order. 
 Noting that the wave vectors correspond to spatial gradients $\nabla$ in real space 
($\bm q$ acting on the electric field, and $\bm Q$ acting on the impurity concentration), 
the spin accumulation (\ref{spin_tot}) and spin current (\ref{sc_tot}) are expressed in real space as 
\begin{widetext}
\begin{align}
 \average{\hat {\bm \sigma}(\bm r)} 
&= 
-\frac{\alpha_{\text{SH}}}{-e} \hat \Pi_{\text s}(\nabla ) \Bigl[ 1+ \nabla \cdot (\delta D(\bm r) \nabla ) \hat \Pi_{\text s}(\nabla ) \Bigr]
 \nabla \times \average{\hat {\bm j}_{\text e}(\bm r)}_0 
- \hat \Pi_{\text s}(\nabla ) \frac{\alpha^{\text{ss}}_{\text{SH}}}{-e} 
   \nabla \times \average{\hat {\bm j}_{\text e}(\bm r)}_{\delta} , 
\label{spin_r}
\\
 \average{\hat j^{\alpha}_{\text s, i}(\bm r)} 
&=  
\epsilon_{\alpha ij} \left[ 
\frac{\alpha_{\text{SH}}}{-e} \average{\hat j_{\text e, j} (\bm r)}_0 + \frac{\alpha^{\text{ss}}_{\text{SH}}}{-e} \average{\hat j_{\text e, j}(\bm r)}_{\delta} 
\right]
-\frac{\alpha^{\text{sj}}_{\text{SH}}}{-e} \epsilon_{\alpha ij} D \nabla_j  
  \average{\hat \rho_{\text e} (\bm r)}_{\delta} 
-D(\bm r) \nabla_i \average{\hat \sigma^{\alpha}(\bm r)}_0
-D\nabla_i\average{\hat \sigma^{\alpha}(\bm r)}_{\delta} 
\label{sc_r}
,
\end{align}
\end{widetext}
where $\hat \Pi_{\text s}(\nabla ) = (-D\nabla^2 +\partial_t + \tau_{\text{sf}}^{-1})^{-1}$ 
is the spin diffusion propagator expressed in real space and $D(\bm r) = D + \delta D(\bm r)$ 
is the diffusion coefficient that includes the effects of inhomogeneity, 
$\delta D(\bm r) = \delta D(\bm Q) \, e^{i\bm Q\cdot \bm r}$.

 In the following, we drop the brackets $\average{\cdots}$ and simply write as 
${\bm \sigma} ({\bm r}) = \average{\hat {\bm \sigma} ({\bm r})}$, etc. 
 Quantities with explicit dependence on ${\bm r}$ include the spatially-modulated parts 
up to the first order in $\delta n_{\rm i} ({\bm r})$. 
 The spin accumulation (\ref{spin_r}) and spin current (\ref{sc_r}) 
satisfy the generalized spin diffusion equation, 
\begin{gather}
 \Bigl\{
\nabla \cdot [D(\bm r) \nabla ] - \partial_t - \tau_{\text{sf}}^{-1}
\Bigr\}
\sigma^{\alpha} (\bm r)
=
\nabla \cdot \bm j_{\text{SH}}^{\alpha} (\bm r)
,
\label{spin_diff_1}
\end{gather}
where $\bm j_{\text{SH}}^{\alpha} (\bm r)$ is the spin Hall current, 
\begin{align}
  j_{\text{SH}, i}^{\alpha} (\bm r) 
= \frac{\alpha_{\text{SH}} ({\bm r})}{-e} \epsilon_{\alpha ij} j_{\text e, j}(\bm r) ,  
\label{eq:j_SH}
\end{align}
and $\bm j_{\text e}(\bm r)$ is the charge current, 
\begin{gather}
  \bm j_{\text e} (\bm r) 
= \sigma_{\text c}(\bm r) \bm E(\bm r) - D(\bm r) \nabla \rho_{\text e}(\bm r) , 
\label{cc_tot}
\end{gather}
which of course satisfies the continuity equation, 
$\partial_t \rho_{\text e}(\bm r) + \nabla \cdot \bm j_{\text e}(\bm r) =0$. 
 The right-hand side of Eq.~(\ref{spin_diff_1}) is the spin source term coming from the divergence 
of the spin Hall current.
 This can be written in the form of ``effective SVC" only if the SH angle is constant 
without spatial modulation, 
$\nabla \cdot \bm j_{\text{SH}}^{\alpha} (\bm r) 
 = \alpha_{\text{SH}} [\nabla \times \bm j_{\text e} (\bm r) ]^\alpha /(-e)$. 

 Finally, if we define the total spin current by the sum of the drift and diffusion spin currents, 
\begin{gather}
 \bm j^{\alpha}_{\text s}(\bm r) 
= \bm j_{\text{SH}}^{\alpha}(\bm r) - D(\bm r) \nabla  \sigma^{\alpha}(\bm r) ,
\label{sc_diff_1}
\end{gather}
Eq.~(\ref{spin_diff_1}) reduces to the spin continuity equation, 
\begin{align}
 \partial_t \sigma^{\alpha} (\bm r) + \nabla \cdot \bm j^{\alpha}_{\text s} (\bm r) 
= -\frac{\sigma^{\alpha}(\bm r)}{\tau_{\text{sf}}} . 
\end{align}

\subsection{Explicit ${\bm Q}$ dependence}

 The explicit dependence on ${\bm Q}$, ${\bm q}$, and $\omega$ of the results are presented 
in Appendix \ref{App:results} [see Eqs.~(\ref{cd_0_result})-(\ref{sc_result})]. 
 Here, we focus on the response to a uniform d.c.~field ${\bm E}$ (${\bm q} = {\bm 0}$, $\omega = 0$) 
and look at the ${\bm Q}$ dependence of the modulations. 
 They are given by 
\begin{align}
  \average{ \hat \sigma^{\alpha}}_\delta 
&= \frac{\sigma_{\rm c}}{-e} \frac{\delta n_{\rm i}}{n_{\rm i}} 
     \alpha_{\rm SH}^{\rm ss} \frac{i({\bm Q} \times {\bm E})_\alpha}{DQ^2 + \tau_{\rm s}^{-1}}  , 
\label{eq:sd_Q}
\\
 \average{ \hat j^{\alpha}_{\text s,i}}_\delta  
&= - \frac{\sigma_{\rm c}}{-e} \frac{\delta n_{\rm i}}{n_{\rm i}}  \biggl\{ 
     \epsilon_{\alpha ij} \left[ 
       \alpha_{\rm SH}^{\rm ss} E_j - \alpha_{\rm SH} \frac{Q_j ({\bm Q} \cdot {\bm E})}{Q^2}  \right]  
\nonumber \\
&\hskip 18mm 
  -  \alpha_{\rm SH}^{\rm ss} \frac{Q_i ({\bm Q} \times {\bm E})_\alpha}{Q^2 + \lambda_{\rm sf}^{-2}} 
    \biggr\}  . 
\label{eq:sc_Q}
\end{align}

\begin{figure}[t]
 \begin{center}
  \includegraphics[width=80mm]{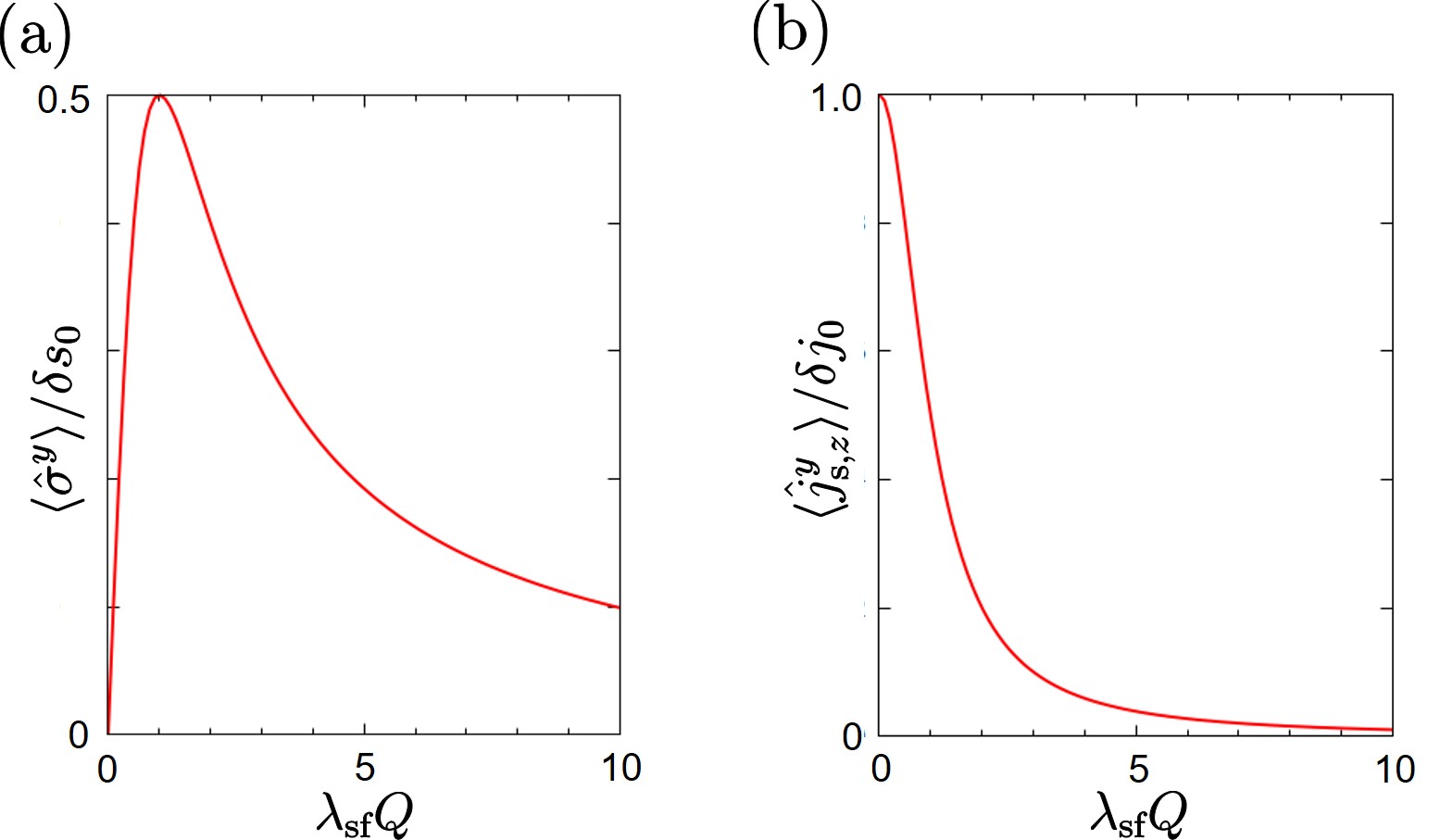}
 \end{center}
 \caption{
 Modulation amplitudes of spin accumulation and spin current induced by a uniform electric field 
${\bm E} = E \hat x$ in the presence of sinusoidal modulation (${\bm Q} = Q \hat z$) 
in the impurity distribution. 
 (a) Spin accumulation $\average{\hat \sigma^y}_\delta$ [Eq.~(\ref{eq:sd_Q})].  
 (b) Spin current $\average{\hat j^y_{\text s, z}}_\delta$ [Eq.~(\ref{eq:sc_Q})].  
 They are normalized by $\delta s_0 = \frac{\lambda_{\rm sf}}{D} \delta j_0$ and 
$\delta j_0 = \alpha_{\rm SH}^{\rm ss} \frac{\delta n_{\rm i}}{n_{\rm i}} \frac{\sigma_{\rm c}}{-e} E$, 
respectively. 
}
 \label{Q}
\end{figure}

 For ${\bm E}$ ($\parallel \hat x$) perpendicular to ${\bm Q}$ ($\parallel \hat z$), 
the spin current modulation is 
\begin{align}
 \average{ j^{\alpha}_{\text s,i}}_\delta  
&= - \frac{\sigma_{\rm c}}{-e} \frac{\delta n_{\rm i}}{n_{\rm i}}  
       \alpha_{\rm SH}^{\rm ss}  \left\{ 
     \epsilon_{\alpha ix} - \delta_{\alpha y} \delta_{iz} \frac{Q^2}{Q^2 + \lambda_{\rm sf}^{-2}}  \right\} E  .  
\end{align}
 The first term is \lq\lq isotropic'' around the current direction, 
as in the ordinary spin Hall current. 
 The second term arises from the spin accumulation [Eq.~(\ref{eq:sd_Q})] by diffusion, 
and is \lq\lq anisotropic''. 
 We plot the $Q$ dependence of 
$\average{\hat \sigma^y }_\delta$ and $\average{\hat j^y_{\text s, z}}_\delta$ in Fig.~\ref{Q}. 
 The former is maximal when $Q^{-1}$ is at the spin relaxation length, 
and the latter simply decreses with $Q$ because of the diffusion spin current 
originating from the former.

 When ${\bm E} $ is parallel to ${\bm Q}$ ($\parallel \hat z$), there arises a charge accumulation, 
$\average{\rho_{\rm e}}_\delta \simeq  \sigma_{\rm c} (\delta n_{\rm i}/n_{\rm i}) (iE/DQ)$, 
and the resulting diffusion charge current induces a spin current modulation, 
\begin{align}
 \average{ j^{\alpha}_{\text s,i}}_\delta  
&=  \frac{\sigma_{\rm c}}{-e} \frac{\delta n_{\rm i}}{n_{\rm i}}   
     \alpha_{\rm SH}^{\rm sj} \epsilon_{\alpha iz}  E    .  
\label{eq:js_E_para_Q}
\end{align}
 However, this is greatly suppressed if the Coulomb interaction is taken into account,  
see Eq.~(\ref{eq:js_E_para_Q_C}).

\subsection{Application}

\begin{figure}[t]
 \begin{center}
  \includegraphics[width=60mm]{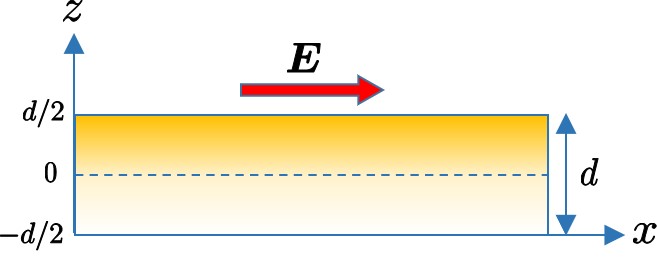}
 \end{center}
 \caption{
Thin film with spin-orbit impurities whose concentration is modulated in the thickness ($z$) direction.} 
 \label{film}
\end{figure}

 As an application of the generalized spin diffusion equation, 
let us consider a thin film of normal metal with spin-orbit impurities whose concentration 
is linearly modulated in the thickness ($z$-) direction (Fig.~\ref{film}), 
\begin{align}
n_{\text i} (z) = n_{\text i} \left[ 1 + \Delta_{\rm i} \frac{z}{d} \right] . 
\end{align}
 The film thickness $d$ is assumed much larger than the electron mean free path. 
 We assume $\Delta_{\rm i}$ is small, and work up to the first order. 
 When a uniform and static external electric field $E$ is applied in the $x$ direction, 
the spin polarization arises in the $y$ direction in both the spin current ($j_{\text s, z}^y$) 
and spin accumulation ($\sigma^y$).

\begin{figure}[h]
 \begin{center}
  \includegraphics[width=88mm]{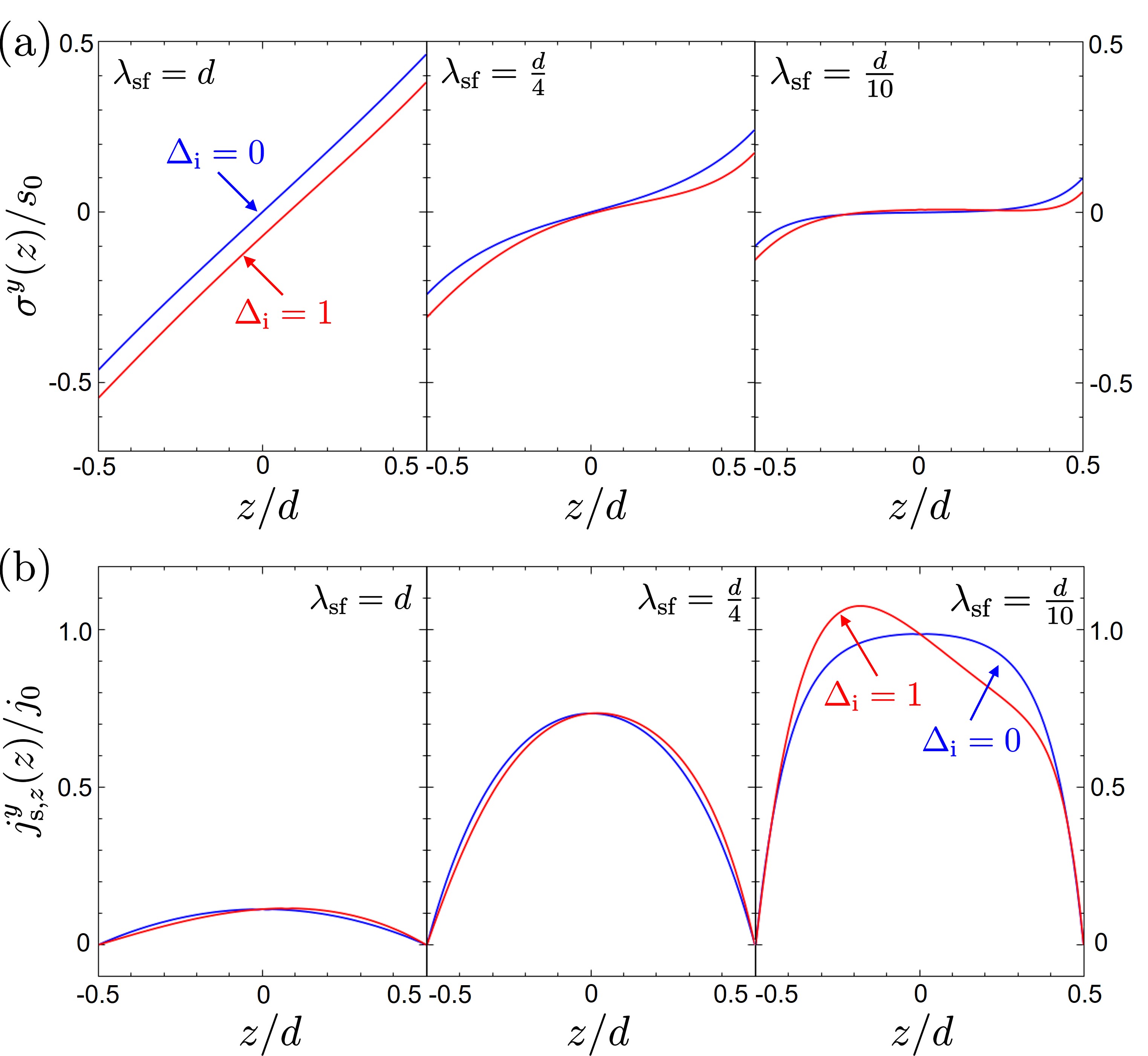}
 \end{center}
 \caption{ 
 Spin Hall effect in a thin film with impurity concentration gradient. 
(a) Spin accumulation $\sigma^y(z)$ normalized by 
$s_0 = \frac{d}{D} \frac{\alpha_{\text{SH}} \sigma_{\text c}}{-e} E$. 
(b) Spin current $j^y_{\text s, z}(z)$ normalized by 
$j_0 = \frac{\alpha_{\text{SH}}\sigma_{\text c}}{-e} E$.
 The blue line is for the homogeneous case ($\Delta_{\text i}=0$), 
and the red line is for the inhomogeneous case ($\Delta_{\text i} \ne 0$). 
 To exaggerate the difference, we set $\Delta_{\rm i}=1$.
 We considered only the skew scattering process, neglecting the side-jump process, 
and put $\alpha^{\rm ss}_{\rm SH} / \alpha_{\rm SH} = 1$. 
}
 \label{plot}
\end{figure}

 We consider the generalized spin diffusion equation,  
\begin{align}
 \partial_z [ D(z) \partial_z \sigma^y (z)] 
- \tau_{\text{sf}}^{-1} (z) \sigma^y (z) = \partial_z j^y_{\text{SH},z} (z) , 
\label{diff_pheno_1}
\end{align}
where $D(z) = D( 1 + \Delta_{\rm i} z/d )^{-1} \simeq D( 1 - \Delta_{\rm i} z/d )$ 
and $\tau_{\rm sf}^{-1} (z) = (\tau_{\rm sf})^{-1} (1 + \Delta_{\rm i} z/d)$ 
are the diffusion constant and the spin-relaxation rate. 
 We have incorporated the effect of $\Delta_{\rm i}$ also in $\tau_{\rm sf}^{-1} $, 
which would result if we went to the third order in $\lambda_{\rm so}$ 
in the microscopic calculation. 
 The right-hand side is the spin source term due to the SH current, 
\begin{align}
 j^y_{\text{SH}, z}(z) 
= \left[ \alpha_{\text{SH}}  -\alpha^{\text{ss}}_{\text{SH}}  \Delta_{\rm i} \frac{z}{d} \right] 
   \frac{\sigma_{\text c}}{-e} E , 
\end{align}
where $\sigma_{\text c}$ is the electrical conductivity at $z=0$. 
 The spin current flowing in the $z$ direction is 
\begin{align}
 j_{\text s, z}^y (z) =  j^y_{\text{SH}, z}(z) - D(z) \partial_z  \sigma^y (z) . 
\end{align}
 As boundary conditions, we impose that no spin currents flow across the surfaces,
$\bm j_{\text s, z} (\frac{d}{2}) = \bm j_{\text s, z}(-\frac{d}{2})=\bm 0$.
 The spin accumulation and spin current are then obtained as 
\begin{widetext}
\begin{align}
 \sigma^y (z) 
&=  s_0 \frac{\lambda_{\text{sf}}}{d}  
  \biggl\{ 
  \frac{\sinh \left( z/\lambda_{\text{sf}} \right) }{\cosh \left( {d}/{2\lambda_{\text{sf}}} \right)} 
 + \Delta_{\rm i} \biggl[ 
   \frac{ \alpha^{\text{ss}}_{\text{SH}} }{\alpha_{\text{SH}}}  
 \left( \frac{\lambda_{\text{sf}}}{d} 
- \frac{\cosh \left( {z}/{\lambda_{\text{sf}}} \right)}{2\sinh \left( {d}/{2\lambda_{\text{sf}}} \right)}
\right) 
 + \frac{z^2 - (d/2)^2}{2d \lambda_{\rm sf}}  
  \frac{\cosh (z/\lambda_{\rm sf})}{\cosh \left( {d}/{2\lambda_{\text{sf}}} \right)} 
 \biggr] \biggr\} ,
\label{spin_pheno}
\\
  j_{\text s, z}^y (z) 
&=  j_0   \biggl\{ 
 \left( 1 - \frac{\cosh \left( {z}/{\lambda_{\text{sf}}} \right)}{\cosh \left( {d}/{2\lambda_{\text{sf}}} \right) } \right) 
 - \Delta_{\rm i} \biggl[ 
   \frac{ \alpha^{\text{ss}}_{\text{SH}} }{\alpha_{\text{SH}}}  
 \left( \frac{z}{d} 
  - \frac{\sinh \left( {z}/{\lambda_{\text{sf}}} \right)}{2\sinh \left( {d}/{2\lambda_{\text{sf}}} \right)}
    \right) 
 + \frac{z^2 - (d/2)^2}{2d \lambda_{\rm sf}}  
  \frac{\sinh (z/\lambda_{\rm sf})}{\cosh \left( {d}/{2\lambda_{\text{sf}}} \right)} 
\biggr] 
   \biggr\}  ,
\label{sc_pheno}
\end{align}
\end{widetext}
to the first order in $\Delta_{\rm i}$, 
where $j_0 = \frac{\alpha_{\text{SH}}\sigma_{\text c}}{-e} E$, $s_0 = \frac{d}{D} j_0$, 
and $\lambda_{\text{sf}}=\sqrt{D\tau_{\text{sf}}}$ is the spin diffusion length.  
 In $\sigma^y$ ($j_{\text s, z}^y$), the contribution for the homogeneous case 
[$\sim O(\Delta_{\rm i}^0)$] is an odd (even) function of $z$, 
and the corrections due the inhomogeneity [$\sim O(\Delta_{\rm i}^1)$] is even (odd).

 The normalized spin accumulation and spin current are plotted in Figs.~\ref{plot} (a) and (b), 
respectively, as functions of $z$ for several choices of spin diffusion length $\lambda_{\rm sf}$.
 As seen, the impurity inhomogeneity enhances the spin accumulation at the $z=-\frac{d}{2}$ 
surface compared to the homogeneous system. 
  This reflects the effects of $j_{\rm SH}$- and $\tau_{\rm sf}^{-1}$-modulations, 
partly canceled by the opposite behavior due to the $D$-modulation. 
 The spin current is suppressed at the surfaces $z = \pm d/2$ because of the boundary condition, 
and takes a maximum inside, whose value develops as $\lambda_{\rm sf}$ is reduced. 
 The effect of $\Delta_{\rm i}$ is opposite between the case $\lambda_{\rm sf} = d, d/4$, 
and the case $\lambda_{\rm sf} = d/10$. 
 The former reflects the effects of $D$- and $\tau_{\rm sf}^{-1}$-modulations, 
and the latter reflects the effects of $j_{\rm SH}$-modulation.

\section{Inclusion of Coulomb interaction}
\label{with}

 The nonequilibrium processes we studied in the preceding sections involve electron 
charge accumulation. 
 See Eq.~(\ref{eq:js^y_Q}), for example, which contains $\average{\hat \rho_{\text e}}_{\delta}$. 
 Also, the impurity inhomogeneity is likely to induce electron charge inhomogeneity even at equilibrium. 
 In metals, however, such charge accumulation and inhomogeneity are screened by other electrons 
at long scales. 
 To discuss these effects properly, it is necessary to consider the long-range Coulomb interaction.

 In this section, we add the Coulomb interaction between electrons, 
$H_{\text C}$ [Eq.~(\ref{eq:Coulomb})], 
to the model $H$ [Eq.~(\ref{H})], and consider $H + H_{\text C}$. 
 We first study the self-energy in the Hartree approximation 
and discuss the electron density in the equilibrium state (Sec.~\ref{C_self}). 
 In Sec.~\ref{C_soc}, we argue that the SOI is not screened, 
and then study in Sec.~\ref{C_response} the response functions by treating $H_{\text C}$ 
in the random phase approximation (RPA). 
 We briefly consider the case of charged impurities in Sec.~\ref{C_charged}, 
and summarize this section in Sec.~\ref{C_summary}.

\begin{figure}[t]
 \begin{center}
  \includegraphics[clip,width=70mm]{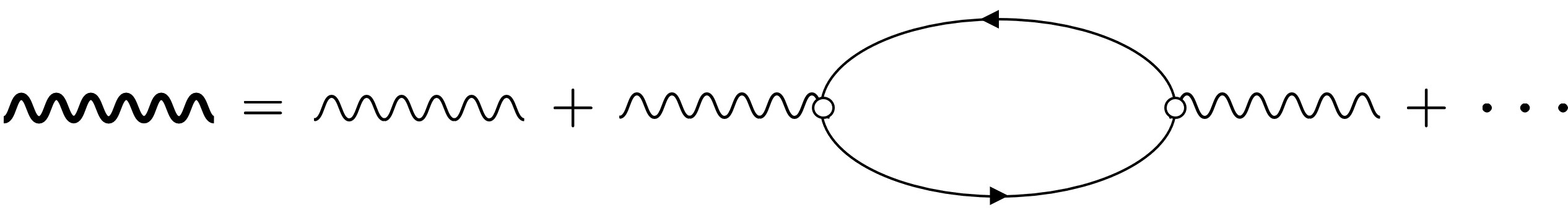}
  \caption{
Effective Coulomb interaction in RPA. 
  The thin (thick) wavy line represents the bare (effective) Coulomb interaction.}
  \label{v_eff}
 \end{center}
\end{figure}
\begin{figure}[t]
 \begin{center}
  \includegraphics[clip,width=70mm]{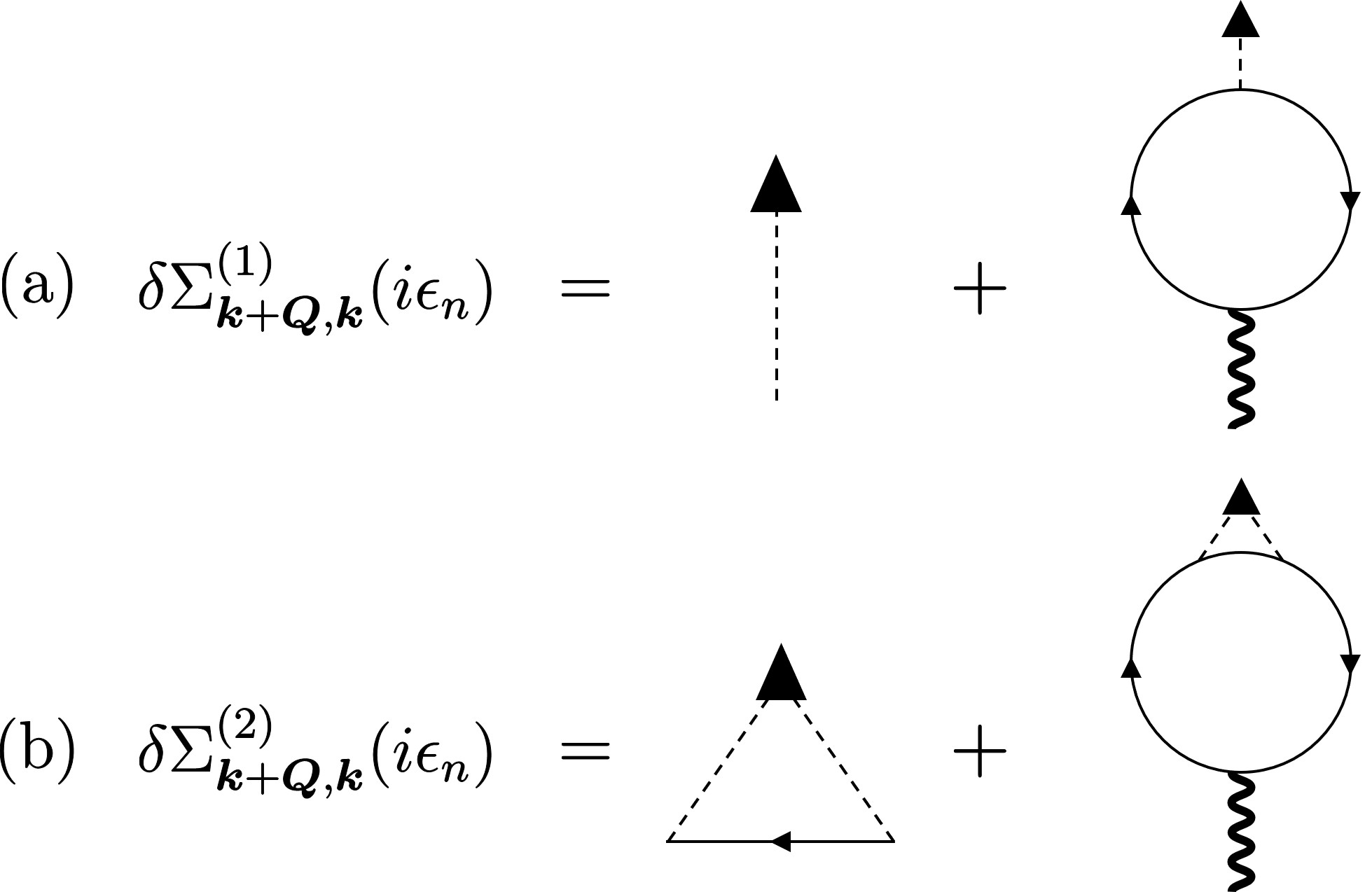}
  \caption{
 Electron self-energy $\delta \Sigma^{(n)}$ at the first order in $\delta n_{\rm i}$ 
and $n$-th order in $u_{\rm i}$ in the presence of  long-range Coulomb interaction. 
 (a) $\delta \Sigma^{(1)}$. (b) $\delta \Sigma^{(2)}$. 
 These are off-diagonal elements with respect to the wave-vector, 
and their nonzero values mean electron density modulation. 
   In $\delta \Sigma^{(1)}$, the first term is given by $\delta n_{\text i}(\bm Q) u_{\rm i}$, 
   an average potential that makes the electron density inhomogeneous. 
   The resulting charge accumulation produces an electric field (or electrostatic potential), 
whose effects on other electrons are expressed by the second (Hartree-like) term. 
   The second term cancels the first term at long wavelength (${\bm Q} \to {\bm 0}$), 
and this is nothing but the screening effect. 
   The same happens for $\delta \Sigma^{(2)}$ as well. 
   The wavy bold line represents the effective Coulomb interaction $U^{\text{eff}}_{\bm Q}$ 
  given by Eq.~(\ref{v_Q}) [Fig.~\ref{v_eff}] with $i\omega_l = 0$. 
}
  \label{delta_self}
 \end{center}
\end{figure}

\subsection{Electron density in equilibrium state}
\label{C_self}

 The inhomogeneity of the impurity distribution may induce an inhomogeneity 
in the electron density. 
 This is described by the first diagram in Fig.~\ref{delta_self} (a) for the self-energy. 
 However, in metals, any charge is screened by other electrons 
because of the long-range Coulomb interaction. 
 This is described by the second diagram of Fig.~\ref{delta_self} (a).

 At the first order in both $\delta n_{\text i}$ and $u_{\text i}$, 
the electron self-energy is thus given by
\begin{align}
\delta \Sigma^{(1)}_{{\bm k}+{\bm Q}, {\bm k}} (i\varepsilon_n) 
= \delta n_{\text i} (\bm Q) u_{\text i} 
   \Bigl[ 1 - U^{\text{eff}}_{\bm Q} (0) \pi^{\text{cc}}_{00} (\bm Q, i\omega_l = 0) \Bigr] , 
\label{eq:Hartree1}
\end{align}
where $\pi^{\text{cc}}_{00} (\bm Q, i\omega_l)$ is the polarization function 
(density-density response function, see Eq.~(\ref{app:pi^cc})), 
and $U_{\bm Q}^{\text{eff}}$ is the effective potential, 
\begin{align}
 U^{\text{eff}}_{\bm Q} (i\omega_l) 
&= \frac{U_{\bm Q}}{1+U_{\bm Q} \pi^{\text{cc}}_{00} (\bm Q, i\omega_l)}.
\label{v_Q}
\end{align}
 In the small-$Q$ limit, $Q/k_{\text F} \ll 1$, 
$\pi^{\text{cc}}_{00} (\bm Q, 0) = 2 N(\mu) + O(Q^2)$, and 
the static effective interaction reduces to the screened Coulomb interaction, 
\begin{align}
 U^{\text{eff}}_{\bm Q} (0)  &\simeq  \frac{e^2}{\epsilon_0 (Q^2 + q_{\text T}^2)} , 
\label{v_Q_small}
\end{align}
where $q_{\text T}^{-1} = \sqrt{ \epsilon_0/ 2e^2 N(\mu)}$ is the Thomas-Fermi screening length. 
 The off-diagonal part of the self-energy is then 
\begin{align}
 \delta \Sigma^{(1)}_{{\bm k}+{\bm Q}, {\bm k}} &= 0 + O(Q^2) . 
\end{align} 
 Therefore, as far as the terms of $\sim O((Q/k_{\text F})^2)$ are neglected, 
the off-diagonal self-energy vanishes and the electron density is uniform in the equilibrium state. 
 The same applies to the second-order contributions 
($\delta \Sigma^{(2)} \sim u_{\text i}^2 \delta n_{\text i}$) in $u_{\text i}$,  
depicted in Fig.~\ref{delta_self} (b). 
 These facts have been used implicitly in the preceding sections.

 The above results hold for 3D. 
 For 2D, the screened Coulomb interaction is given by 
\begin{align}
 U^{\rm eff}_{\bm Q} (0)  &\simeq  \frac{e^2}{2\epsilon_0 (Q + q_{\text T})} , 
\label{v_Q_small_2D}
\end{align}
with $q_{\text T} = e^2 N(\mu)/\epsilon_0$, 
and the off-diagonal self-energy $\delta \Sigma^{(1)}_{{\bm k}+{\bm Q}, {\bm k}}$ 
vanishes like $\sim O(Q)$. 
 In the following, to avoid complexity, we mainly focus on 3D. 
 Most results for 2D can be obtained from those for 3D by the replacement, 
\begin{align}
 q_{\text T}^2 = \frac{2e^2 N(\mu)}{\epsilon_0}   \ \ {\rm (3D)} \  \to  \ \  
 \frac{e^2 N(\mu)}{\epsilon_0} q'  \equiv q_{\text T} q'  \ \ {\rm (2D)} , 
\label{3D_2D}
\end{align}
where $q'=q$ or $|{\bm q} + {\bm Q}|$.

\subsection{Spin-orbit coupling}
\label{C_soc}

 One may suppose that the electric field (or the potential $V$) that defines the SOI, 
Eq.~(\ref{soi}), is similarly screened by the Coulomb interaction. 
 Here, we argue that this is not necessarily the case. 
 Specifically, we show or discuss the followings.
(i) The SOI is not screened in the present model. 
(ii) If the model is further extended to include relativistic corrections to the Coulomb interaction, 
    the SOI can, in principle, be screened. 
(iii) Screening of SOI occurs if the density of core electrons is changed, 
which is unlikely in the present processes.

(i)  In the present model ($H + H_{\text C}$), the SOI is not screened. 
 This is explicitly seen from the fact that the electron loop of the second diagram 
of Fig.~\ref{delta_self} (a) vanishes if the impurity line comes from SOI [Eq.~(\ref{soi})], 
\begin{align}
  T \sum_n \sum_{\bm k} i\lambda_{\rm so} V_{\bm Q} ({\bm Q} \times {\bm k})  \cdot 
  {\rm tr} [G_{{\bm k}+{\bm Q}} {\bm \sigma}  G_{\bm k}] = 0  . 
\end{align} 
 Note that the Coulomb interaction is associated with a unit vertex, whereas the SOI 
has Pauli matrices, making the spin trace vanish. 
 This means that the electron bubble does not connect to the impurity SOI, 
and thus the SOI is not \lq\lq screened'' in the present model.

(ii) It is possible that the SOI is \lq\lq screened'' if the relativistic corrections 
to the Coulomb interaction are considered. 
 Such interactions are known as the Breit interaction.\cite{Breit} 
 In Appendix \ref{Breit}, we illustrate how the screening of SOI occurs via the Breit interaction.

 (iii) We note that the (appreciable) SOI originates from the strong electric field near the nucleus, 
partly screened by \lq\lq core'' electrons. 
 The length scale that determines the effective SOI  parameter $\lambda_{\rm so}$ in Eq.~(\ref{soi}) 
is thus much shorter than the screening length due to conduction electrons. 
 Therefore, the ordinary metallic screening as considered here 
is not effective to screen the SOI.

 As such, we assume the SOI is not screened in the low-energy processes 
that we consider in this paper. 
 This means that we can proceed with $H + H_{\text C}$.

\subsection{Response functions}
\label{C_response}

 In harmony with the Hartree approximation, 
we use RPA to evaluate the dynamical response functions. 
 As a result, the charge accumulations are suppressed also in the dynamical processes.

 As shown in Appendix \ref{app_coulomb}, 
the spin accumulation and spin-current density are obtained as 
\begin{widetext}
\begin{align}
  \average{\hat \sigma^{\alpha}}^{\text{RPA}} 
&= -  \Pi_{\text s}(\bm q) \frac{\alpha_{\text{SH}}}{-e} [ i\bm q \times  \average{\hat{\bm j}_{\text e}}_0^{\text{RPA}} ]_{\alpha}
\nonumber \\ 
&\ \ 
   - \Pi_{\text s}(\bm q+\bm Q) \biggl\{ \frac{\alpha^{\text{ss}}_{\text{SH}}}{-e} [ i(\bm q +\bm Q) \times    \average{\hat{\bm j}_{\text e}}_{\delta}^{\text{RPA}} ]_{\alpha}
  - \frac{\alpha_{\text{SH}}}{-e}  \delta D(\bm Q) \, 
    \bm q \cdot (\bm q + \bm Q) \Pi_{\text s}(\bm q) 
    [ i\bm q  \times \average{\hat{\bm j}_{\text e}}_0^{\text{RPA}} ]_{\alpha} \biggr\} ,
\label{spin_tot_rpa}
\\
 \average{\hat j^{\alpha}_{\text s,i}}^{\text{RPA}} 
&=  \alpha_{\text{SH}} \epsilon_{\alpha ij}  \frac{\average{j_{\text e, j}}_0^{\text{RPA}} }{-e}
  + \epsilon_{\alpha ij} 
    \biggl[ \frac{\alpha_{\text{SH}}^{\text{ss}}}{-e} \average{ \hat j_{\text e, j}}_{\delta}^{\text{RPA}}
   + \frac{\alpha^{\text{sj}}_{\text{SH}}}{-e}  
    \left\{  \sigma_{\rm c} E_{\delta, j}^{\rm ind}  
            - Di(q_j+Q_j) \average{\hat \rho_{\text e}}_{\delta}^{\text{RPA}} \right\} 
 \biggr]
\nonumber \\ 
& \ \ 
- \left[ D+\delta D (\bm Q) \right] iq_i \average{\hat \sigma^{\alpha}}_0^{\text{RPA}}
-Di(q_i+Q_i) \average{\hat \sigma^{\alpha}}_{\delta}^{\text{RPA}},
\label{sc_tot_rpa}
\end{align}
where the charge and charge-current densities are given by
 \begin{align}
&\average{\hat \rho_{\text e}}^{\text{RPA}}_{\delta} 
=  \sigma_{\text c} \frac{\delta n_{\text i}}{n_{\text i}} 
   \frac{1}{D[(\bm q+\bm Q)^2 + q_{\text T}^2] -i\omega} 
   \biggl[ i(\bm q+\bm Q) \cdot \left( \bm E + \bm E^{\text{ind}}_0   \right) 
          - \frac{D \, \bm q \!\cdot\! (\bm q+\bm Q) i\bm q\cdot \bm E}{D(q^2+q_{\text T}^2)-i\omega} \biggr] ,
\label{eq:rho_delta_RPA}
\\
&\average{\hat{\bm j_{\text e}}}^{\text{RPA}}_{\delta} 
=  \sigma_{\text c} 
 \left[ - \frac{\delta n_{\text i}}{n_{\text i}} 
         \left( \bm E+\bm E^{\text{ind}}_0  \right) +  \bm E^{\text{ind}}_\delta   \right] 
 - \delta D(\bm Q) i\bm q \average{\hat \rho_{\text e}}_0^{\text{RPA}}
 - D  i(\bm q + \bm Q) \average{\hat \rho_{\text e}}^{\text{RPA}}_{\delta}  , 
\end{align}
\end{widetext}
at first order in $\delta n_{\text i}$, and they are given by
\begin{align}
 \average{\hat \rho_{\text e}}^{\text{RPA}}_0 
&= -\sigma_{\text c} \frac{i\bm q \cdot \bm E}{D(q^2+q_{\text T}^2) -i\omega},
\label{eq:rho_0_RPA}
\\
 \average{\hat{\bm j}_{\text e}}_0^{\text{RPA}} 
&=  \sigma_{\text c}\left( \bm E + \bm E^{\text{ind}}_0  \right) 
    - Di\bm q \, \average{\hat \rho_{\text e}}^{\text{RPA}}_0 .
\end{align}
at zeroth order. 
 We defined the electric fields, 
\begin{align}
 \bm E^{\text{ind}}_0  &= -i\bm q \frac{\average{\hat \rho_{\text e}}_0^{\text{RPA}}}{\epsilon_0 q^2},
\label{eq:E^ind_0}
\\
 \bm E^{\text{ind}}_\delta  
&= -i(\bm q + \bm Q) 
  \frac{\average{\hat \rho_{\text e}}^{\text{RPA}}_{\delta}}{\epsilon_0 (\bm q+\bm Q)^2} , 
\label{eq:E^ind_delta}
\end{align}
produced by the ${\bm E}$-induced charge accumulations, 
$\average{\hat \rho_{\text e}}^{\text{RPA}}_0$ and $\average{\hat \rho_{\text e}}^{\text{RPA}}_{\delta}$, 
respectively. 
 [The real-space form is given by Eq.~(\ref{eq:Eind_real}) below.] 
 In Eq.~(\ref{sc_tot_rpa}), 
 $\average{\hat \sigma^{\alpha}}_{0}^{\text{RPA}}$ represents the first term in Eq.~(\ref{spin_tot_rpa}), 
and $\average{\hat \sigma^{\alpha}}_{\delta}^{\text{RPA}}$ the remaining terms. 
 Therefore, the Coulomb interaction modifies the results in two ways. 
 First, it gives a \lq\lq mass'' to the charge diffusion propagator, 
$(Dq^2 -i\omega)^{-1} \to [D(q^2+q_{\text T}^2) -i\omega]^{-1} \equiv \Pi_{\rm c}$.  
 (The spin diffusion propagator $\Pi_{\text s}({\bm q}) = (Dq^2 - i\omega + \tau_{\text{sf}}^{-1})^{-1}$ 
is not modified.) 
 Second, it induces additional (screening) fields, 
$\bm E^{\text{ind}}_0$ and $\bm E^{\text{ind}}_\delta$.

 In metals, $q_{\rm T}$ is comparable with the inverse interatomic distance, 
hence $q \ll q_{\rm T}$ and $\omega \ll Dq_{\rm T}^2$ hold well. 
 Therefore, the charge diffusion propagator is well approximated as 
$\Pi_{\rm c} \simeq (Dq_{\rm T}^2)^{-1} = \epsilon_0 / \sigma_{\rm c}$, 
and Eqs.~(\ref{eq:rho_0_RPA}) and (\ref{eq:rho_delta_RPA}) become  
\begin{align}
 \average{\hat \rho_{\text e}}^{\text{RPA}}_0 &\simeq - \epsilon_0 (i\bm q \cdot \bm E ) , 
\label{eq:rho_0_RPA_approx}
\\
 \average{\hat \rho_{\text e}}^{\text{RPA}}_{\delta} 
&\simeq  \epsilon_0 \frac{\delta n_{\text i}}{n_{\text i}} 
    i(\bm q+\bm Q) \cdot \left( \bm E + \bm E^{\text{ind}}_0   \right) . 
\label{eq:rho_delta_RPA_approx}
\end{align}
 From Eqs.~(\ref{eq:E^ind_0}) and (\ref{eq:rho_0_RPA_approx}), one has 
${\bm E}^{\text{ind}}_0 = - {\bm q} ({\bm q} \cdot {\bm E})/q^2 \equiv - {\bm E}_\parallel$, 
meaning that 
$\average{\hat \rho_{\text e}}^{\text{RPA}}_0$ completely screens the longitudinal component  
${\bm E}_\parallel$ of the applied electric field ${\bm E}$. 
 Using this result in Eq.~(\ref{eq:rho_delta_RPA_approx}), we have 
$\average{\hat \rho_{\text e}}^{\text{RPA}}_{\delta} 
 \simeq  \epsilon_0 \frac{\delta n_{\text i}}{n_{\text i}} i({\bm Q} \cdot  {\bm E}_\perp ) 
= \epsilon_0 \frac{1}{n_{\text i}} ({\bm \nabla} n_{\text i} \cdot  {\bm E}_\perp ) 
=  - \epsilon_0 
   \frac{1}{\sigma_{\rm c}} {\rm div} (\sigma_{\rm c} ({\bm r})  {\bm E}_\perp ({\bm r})) $, 
where ${\bm E}_\perp = {\bm E} - {\bm E}_\parallel$ is the transverse component. 
 Then the total induced field, ${\bm E}^{\rm ind} = {\bm E}^{\rm ind}_0 + {\bm E}^{\rm ind}_\delta$,  
is given by 
\begin{align}
 {\bm E}^{\rm ind} ({\bm r}) 
&= - {\bm E}_\parallel ({\bm r}) 
    - \frac{1}{\sigma_{\rm c}} [\delta \sigma_{\rm c} ({\bm r})  {\bm E}_\perp ({\bm r})]_\parallel  .  
\label{eq:E^ind}
\end{align}
 The second term represents the screening by 
$\langle \hat \rho_{\rm e} \rangle^{\rm RPA}_\delta$ 
of the additional longitudinal field, 
$\frac{1}{\sigma_{\rm c}} [\delta \sigma_{\rm c} ({\bm r})  {\bm E}_\perp ({\bm r})]_\parallel$, 
produced by the inhomogeneity $\delta n_{\rm i}$.

 In real space, the total spin accumulation and the total spin-current density are expressed as
\begin{widetext}
\begin{align}
 \average{\hat {\bm \sigma}(\bm r)}^{\text{RPA}}
&= 
-\frac{\alpha_{\text{SH}}}{-e} \hat \Pi_{\text s}(\nabla ) \Bigl[ 1+ \nabla \cdot (\delta D(\bm r) \nabla ) \hat \Pi_{\text s}(\nabla ) \Bigr]
 \nabla \times \average{\hat {\bm j}_{\text e}(\bm r)}_0^{\text{RPA}}
- \hat \Pi_{\text s}(\nabla ) \frac{\alpha^{\text{ss}}_{\text{SH}}}{-e} 
   \nabla \times \average{\hat {\bm j}_{\text e}(\bm r)}_{\delta}^{\text{RPA}}, 
\label{spin_r_rpa}
\\
 \average{\hat j^{\alpha}_{\text s, i}(\bm r)}^{\text{RPA}}
&=  
\epsilon_{\alpha ij} \left[ 
\frac{\alpha_{\text{SH}}}{-e} \average{\hat j_{\text e, j} (\bm r)}_0^{\text{RPA}} + \frac{\alpha^{\text{ss}}_{\text{SH}}}{-e}
 \average{\hat j_{\text e, j}(\bm r)}_{\delta}^{\text{RPA}}
\right]
-\frac{\alpha^{\text{sj}}_{\text{SH}}}{-e} \epsilon_{\alpha ij} D \nabla_j  
  \average{\hat \rho_{\text e} (\bm r)}_{\delta}^{\text{RPA}}
  \nonumber  \\ &
-D(\bm r) \nabla_i \average{\hat \sigma^{\alpha}(\bm r)}_0^{\text{RPA}}
-D\nabla_i\average{\hat \sigma^{\alpha}(\bm r)}_{\delta}^{\text{RPA}} .
\label{sc_r_rpa}
\end{align}
\end{widetext}
 They satisfy the generalized spin diffusion equation (\ref{spin_diff_1}), 
with the spin Hall current ${\bm j}^{\alpha}_{\text{SH}} (\bm r)$ having the same form as Eq.~(\ref{eq:j_SH}), 
but the electric current ${\bm j}_{\rm e}({\bm r})$ in it is now updated to 
\begin{align}
 \bm j_{\text e}(\bm r) 
&= \sigma_{\text c}(\bm r) \Bigl( \bm E(\bm r) + \bm E^{\text{ind}}(\bm r) \Bigr) 
                               - D(\bm r) \nabla \rho_{\text e} (\bm r) 
\label{cc_rpa_00}
\\
& \simeq [ \, \sigma_{\text c} ({\bm r}) {\bm E}_\perp ({\bm r})]_\perp      ,
\label{cc_rpa}
\end{align}
where 
\begin{align}
 \bm E^{\text{ind}}(\bm r) 
= - \nabla \int d^3r'   \frac{1}{4\pi \epsilon_0} \frac{ \rho_{\text e}(\bm r')}{|\bm r-\bm r'|} , 
\label{eq:Eind_real}
\end{align} 
is the electric field produced by the induced (total) charge accumulation, 
$\rho_{\rm e}({\bm r}) 
 = \average{\hat \rho_{\rm e}}^{\rm RPA}_0 + \average{\hat \rho_{\rm e}}^{\rm RPA}_\delta$. 
 In Eq.~(\ref{cc_rpa_00}), we used Eq.~(\ref{eq:E^ind}) and neglected the diffusion current, 
which is smaller by $\sim (q/q_{\rm T})^2$ than the drift current. 
 The result indicates that the current induced by the transverse electric field 
${\bm E}_\perp$ (which survived the screening by $\langle \hat \rho_{\rm e} \rangle^{\rm RPA}_0$) 
can have longitudinal component in $\delta n_{\rm i} ({\bm r}) {\bm E}_\perp$ 
because of the ${\bm r}$-dependence of $\delta n_{\rm i}$. 
 The associated charge accumulation $\langle \hat \rho_{\rm e} \rangle^{\rm RPA}_\delta$ 
is further screened (the electric field further subtracted by the last term in Eq.~(\ref{eq:E^ind})), 
resulting in a purely transverse (divergenceless) current. 
 This is consistent with the time-independence of the electrical charge density, 
as required by the charge neutrality \lq\lq constraint'' in metals.

\subsection{Explicit ${\bm Q}$ dependence} 

 Let us focus on the response to a uniform d.c.~field ${\bm E}$ (${\bm q} = {\bm 0}$, $\omega = 0$) 
and look at the ${\bm Q}$ dependence. 
 Compared to the results in the preceding section, 
the Coulomb interaction primarily modifies the charge accumulation, 
\begin{align}
&\average{ \rho_{\rm e}}^{\rm RPA}_\delta
=  \sigma_{\rm c} \frac{\delta n_{\rm i}}{n_{\rm i}} 
   \frac{i{\bm Q} \cdot {\bm E}}{D(Q^2 + q_{\text T}^2) } , 
\end{align}
and then the resulting diffusion current and the induced spin current. 
 Thus the spin current modulation becomes 
\begin{align}
 \average{ j^{\alpha}_{\text s,i}}^{\rm RPA}_\delta  
&= - \frac{\sigma_{\rm c}}{-e} \frac{\delta n_{\rm i}}{n_{\rm i}}  \biggl\{ 
       \alpha_{\rm SH}^{\rm ss} \epsilon_{\alpha ij}  \left( E_j  - \frac{Q_j ({\bm Q} \cdot {\bm E})}{Q^2} \right)  
\nonumber \\
& - \alpha_{\rm SH}^{\rm sj} \epsilon_{\alpha ij}  \frac{Q_j ({\bm Q} \cdot {\bm E})}{Q^2 + q_{\rm T}^2}  
      -  \alpha_{\rm SH}^{\rm ss} \frac{Q_i ({\bm Q} \times {\bm E})_\alpha}{Q^2 + \lambda_{\rm sf}^{-2}} 
    \biggr\} .  
\end{align}
 The Coulomb interaction modifies the spin current modulation 
when ${\bm E}$ is parallel to ${\bm Q}$ ($\parallel \hat z$),
\begin{align}
 \average{ j^{\alpha}_{\text s,i}}^{\rm RPA}_\delta  
&= \frac{\sigma_{\rm c}}{-e} \frac{\delta n_{\rm i}}{n_{\rm i}}  
     \alpha_{\rm SH}^{\rm sj} \epsilon_{\alpha iz}  \frac{Q^2}{Q^2 + q_{\rm T}^2} E .  
\label{eq:js_E_para_Q_C}
\end{align}
 Since $Q \ll q_{\rm T}$, this is greatly supressed compared to Eq.~(\ref{eq:js_E_para_Q}).

\subsection{Case of charged impurities} 
\label{C_charged}

 So far, we assumed short-range impurity potential 
appropriate for uncharged (or already screened) impurities.  
 Here we briefly study the case of charged impurities. 

 For simplicity, we consider a monovalent ion for the impurity. 
 They are accounted for by replacing the (point-like) impurity potential $u_{\text i} $ 
by the Coulomb potential, $-U_{\bm K} = -\frac{e^2}{\epsilon_0 K^2}$. 
 With the same distribution as Eq.~(\ref{eq:p_imp}), the off-diagonal self-energy (\ref{eq:Hartree1}) 
is now 
\begin{align}
\delta \Sigma^{(1)}_{{\bm k}+{\bm Q}, {\bm k}} (i\varepsilon_n) 
&= \delta n_{\text i} (\bm Q) (-U_{\bm Q} )
   \Bigl[ 1 - U^{\text{eff}}_{\bm Q} (0) \pi^{\text{cc}}_{00} (\bm Q, 0) \Bigr] 
\nonumber \\
&=  - \frac{\delta n_{\text i} }{2N(\mu)} + O(Q^2) . 
\label{eq:Sigma_charged}
\end{align}
 This remains finite in the limit, ${\bm Q} \to {\bm 0}$, 
meaning that the electron density becomes inhomogeneous. 
 In fact, one has $\delta \rho_{\rm e} ({\bm r}) = \delta n_{\text i} ({\bm r})$, 
and the electrostatic potential produced by this modulation completely 
cancels the original one (due to charged impurities). 
 The result is that the impurities are screened and the electron density is nonuniform.

 Inserting the self-energy correction, Eq.~(\ref{eq:Sigma_charged}), to the ordinary SHE calculation 
(skew-scattering and side-jump processes without $\delta n_{\text i}$), 
we observed that the contributions are higher order in the damping $\tau^{-1}$, 
hence can be neglected. 
  Therefore, the electron charge inhomogeneity at equilibrium does not affect the spin Hall effect. 
 This is of course consistent with the results we obtained 
for uncharged (or already screened) impurities.

\subsection{Short summary} 
\label{C_summary}

 The effects of long-range Coulomb interaction studied in this section are summarized as follows.

 For uncharged impurities, the electron density at equilibrium is kept uniform 
even when the impurity distribution is inhomogeneous. 
 Charge accumulation in the dynamical response process, 
namely, the spin Hall current from the charge accumulation, is also suppressed.

 If the impurities are charged, their inhomogeneity induces electron density inhomogeneity 
(because of screening) at equilibrium. 
 The spin Hall response is not affected by this electron density inhomogeneity, 
and it is quite similar to the case of uncharged impurities.

\section{ Conclusion } 
We studied extrinsic spin Hall effect in systems with inhomogeneously distributed impurities.
We found that the spin accumulation is induced by the inhomogeneity 
via the side-jump and skew-scattering processes. 
 The results satisfy a generalized spin diffusion equation with a spin source term, 
which is expressed as the divergence of spin Hall current 
and reduces to the  ``effective SVC in the laboratory (inertial) frame" if the SH angle is homogeneous. 
 These features are preserved even when the long-range Coulomb interaction is considered,
which suppresses the electron charge accumulation and inhomogeneity.  
 It would be interesting to apply the obtained spin diffusion equation 
to the experiments of spin-current injection from surface oxidized copper\cite{ando1, nitta, nozaki1}. 
 These are left as future studies.

\begin{acknowledgements}
We gratefully acknowledge helpful discussions with A. Yamakage, K. Nakazawa, T. Yamaguchi, Y. Imai, J. J. Nakane, and Y. Yamazaki.
 This work is supported by JSPS KAKENHI Grant Numbers JP15H05702, JP17H02929  
and JP19K03744.
TF is supported by Grant-in-Aid for JSPS Fellows Grant Number 19J15369,
and by a Program for Leading Graduate Schools ``Integrative Graduate Education and Research in Green Natural Sciences''.
\end{acknowledgements}

\appendix

\section{Ladder vertex corrections}

\begin{figure}[t]
 \begin{center}
  \includegraphics[width=75mm]{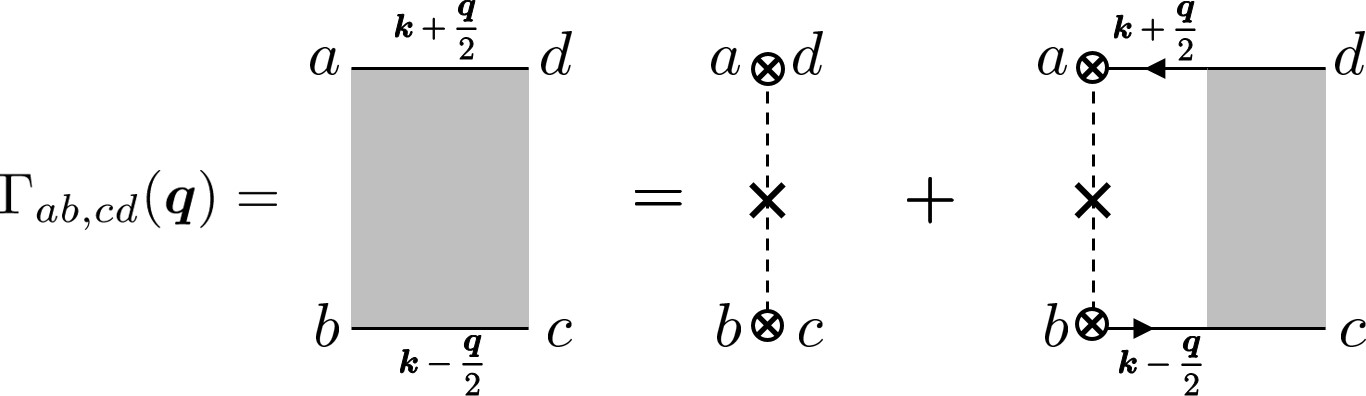}
  \caption{Four-point vertex in the ladder approximation due to homogeneously distributed impurities.
The circle with a cross represents the sum of spin-flip and spin-nonflip vertices.
$a$-$d$ represent spin indices.}
  \label{vcs}
 \end{center}
\end{figure}

 In this Appendix, we calculate ladder vertex corrections due to impurities. 
 The four-point vertex $\Gamma_{ab, cd}$ in the absence of impurity inhomogeneity 
(zeroth order in $\delta n_{\rm i}$), shown in Fig.~\ref{vcs}, is given by 
\begin{align}
\Gamma_{ab, cd}(\bm q) = \Gamma_{\text c}(\bm q) \delta_{ab} \delta_{cd} + \Gamma_{\text s}(\bm q) \bm \sigma_{ab} \cdot \bm \sigma_{cd},
\end{align}
where $a$-$d$ are spin indices, and
\begin{align}
\Gamma_{\text c}(\bm q) &= \frac{1}{4\pi N(\mu ) \tau^2} \frac{1}{Dq^2-i\omega }, \\
\Gamma_{\text s}(\bm q) &= \frac{1}{4\pi N(\mu ) \tau^2} \frac{1-\frac{\tau}{\tau_{\text{sf}}}}{Dq^2-i\omega  + \tau_{\text{sf}}^{-1}}.
\end{align}
 The one in the first order in $\delta n_{\rm i}$, shown in Fig.~\ref{vcs_in} is given by 
\begin{align}
\delta \Gamma_{ab, cd}(\bm Q) = \delta \Gamma_{\text c}(\bm Q)\delta_{ab}\delta_{cd} + \delta \Gamma_{\text s}(\bm Q) \bm \sigma_{ab}\cdot \bm \sigma_{cd},
\end{align} 
where 
\begin{align}
\delta \Gamma_{\rm c(s)} (\bm Q) &=4\Gamma_{\text c(s)}(\bm q+\bm Q) \Gamma_{\text c(s)}(\bm q) 
\frac{\delta n_{\text i} (\bm Q)}{n_{\text i}}\frac{\gamma}{\pi N(\mu )}
\nonumber
\\
&
\times
\Bigl[
I_0(\bm q+\bm Q) I_0(\bm q) -Y_0(\bm Q)  
\Bigr] , 
\end{align}
and $I_0$ and $Y_0$ are given in Appendix~\ref{int}. 
  Thus, 
\begin{align}
\delta \Gamma_{\text c} (\bm Q) 
&=
\frac{\delta n_{\text i}(\bm Q)}{n_{\text i}} \frac{\gamma}{\pi N(\mu )}
\frac{1}{D(\bm q+\bm Q)^2-i\omega}
\frac{D\bm q\cdot (\bm q+\bm Q)}{\tau (Dq^2-i\omega )},
\\ 
\delta \Gamma_{\text s}(\bm Q) &=
\frac{\delta n_{\text i}(\bm Q)}{n_{\text i}} \frac{\gamma}{\pi N(\mu )}
\frac{(1-\frac{\tau}{\tau_{\text{sf}}})^2}{D(\bm q+\bm Q)^2-i\omega + \tau_{\text{sf}}^{-1}}
\nonumber
\\
&
\times 
\frac{D\bm q\cdot (\bm q+\bm Q)}{\tau (Dq^2-i\omega ) + \frac{\tau}{\tau_{\text{sf}}}}.
\end{align}

Next, the three-point vertices, $\Lambda^{\text c}$ and $\Lambda^{\text s}$, 
shown in Fig.~\ref{3vc_uni}, are calculated as
\begin{align}
\delta_{ab}\Lambda^{\text c}_{\nu }(\bm q) &=\Gamma_{ab, cd}(\bm q) \delta_{dc} I_{\nu}(\bm q) ,
\nonumber
\\
&= \delta_{ab}
\begin{cases}
 \dfrac{1}{\tau (Dq^2-i\omega )}-1   &(\nu = 0),
\\
-  \dfrac{Diq_j}{\tau (Dq^2-i\omega )}  &(\nu = j),
\end{cases}
\\
\sigma^{\alpha}_{ab}\Lambda^{\text s}_{\nu }(\bm q) &=\sigma^{\alpha}_{dc} \Gamma_{ab, cd}(\bm q)  I_{\nu} (\bm q) ,
\nonumber
\\
&= \sigma^{\alpha}_{ab}
\begin{cases}
  \dfrac{1}{\tau (Dq^2 -i\omega ) + \frac{\tau}{\tau_{\text{sf}}}} -1
&(\nu =0) ,
\\
- \dfrac{1-\frac{\tau}{\tau_{\text{sf}}}}{\tau (Dq^2 -i\omega ) + \frac{\tau}{\tau_{\text{sf}}}} Diq_j
&(\nu = j).
\end{cases}
\end{align}

\begin{widetext}

\begin{figure}[bp]
 \begin{center}
  \includegraphics[width=120mm]{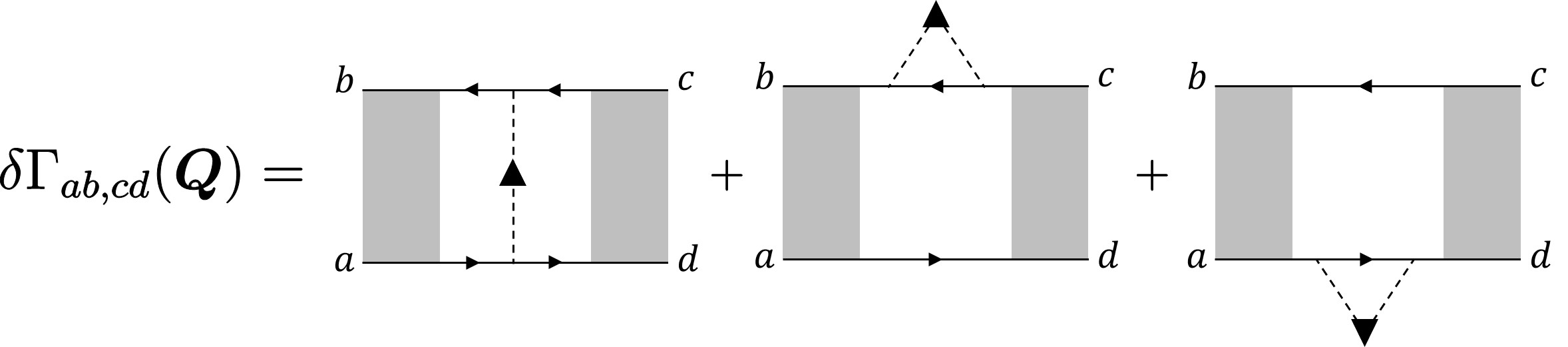}
  \caption{Four-point vertex in the ladder approximation which is first order in the inhomogeneity of normal impurity scattering.}
  \label{vcs_in}
 \end{center}
\end{figure}

\begin{figure}[tbp]
 \begin{center}
  \includegraphics[width=90mm]{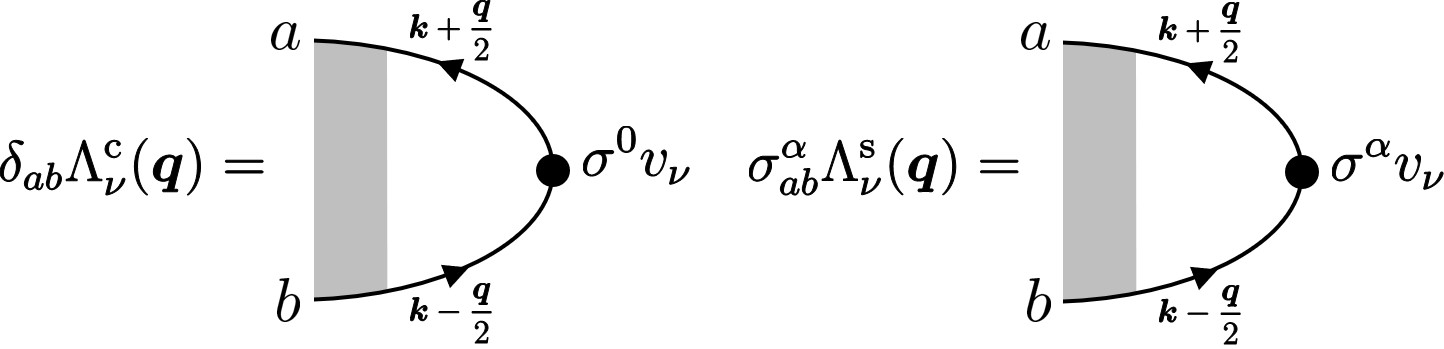}
  \caption{Three-point vertices with ladder vertex corrections of Fig.~\ref{vcs}.}
  \label{3vc_uni}
 \end{center}
\end{figure}

The three-point vertices, $\Pi^{\text{(L)}}$ and $\Pi^{\text{(R)}}$, shown in Fig.~\ref{3vc_lr}, 
are calculated as
\begin{align}
\Pi^{\text{(L)}}_{\mu}(\bm Q) &= 2\Gamma_{\text s} (\bm q) \frac{\delta n_{\text i}(\bm Q)}{n_{\text i}} \frac{\gamma}{\pi N(\mu )} 
\Bigl[
I_0(\bm q) I_{\mu}(\bm q+\bm Q)
-Y_{\mu}(\bm Q) 
\Bigr]
+
\delta \Gamma_{\text s}(\bm Q) I_{\mu}(\bm q+\bm Q)
\\
&=
\begin{cases}
\dfrac{\delta n_{\text i}(\bm Q)}{n_{\text i}} \dfrac{(1-\frac{\tau}{\tau_{\text{sf}}})^2}{D(\bm q+\bm Q)^2-i\omega +\tau_{\text{sf}}^{-1}}
\dfrac{D\bm q\cdot (\bm q+\bm Q)}{\tau (Dq^2-i\omega ) + \frac{\tau}{\tau_{\text{sf}}}}  &  (\mu =0),
\\[10pt]
\dfrac{\delta n_{\text i}(\bm Q)}{n_{\text i}} \left( 1-\dfrac{\tau}{\tau_{\text{sf}}} \right)
\dfrac{Diq_i}{\tau (Dq^2-i\omega )+\frac{\tau}{\tau_{\text{sf}}}}
- Di(q_i+Q_i) \Pi^{\text{(L)}}_{0}(\bm Q) & (\mu =i),
\end{cases}
\\
\Pi^{\text{(R)}}_{\nu}(\bm Q) &= 2\Gamma_{\text c} (\bm q+\bm Q) \frac{\delta n_{\text i}(\bm Q)}{n_{\text i}} \frac{\gamma}{\pi N(\mu )} 
\Bigl[
I_0(\bm q+\bm Q) I_{\nu}(\bm q)
-Y_{\nu}(\bm Q)
\Bigr]
+
\delta \Gamma_{\text c}(\bm Q) I_{\nu}(\bm q)
\\
&=
\begin{cases}
\dfrac{\delta n_{\text i}(\bm Q)}{n_{\text i}} \dfrac{1}{D(\bm q+\bm Q)^2-i\omega}
\dfrac{D\bm q\cdot (\bm q+\bm Q)}{\tau (Dq^2-i\omega )}  &  (\nu =0),
\\[10pt]
\dfrac{\delta n_{\text i}(\bm Q)}{n_{\text i}}
\dfrac{Di(q_j+Q_j)}{\tau [D(\bm q+\bm Q)^2-i\omega ]}
- Diq_j \Pi^{\text{(R)}}_{0}(\bm Q) & (\nu =j) . 
\end{cases}
\end{align}

\begin{figure}[tbp]
 \begin{center}
  \includegraphics[width=160mm]{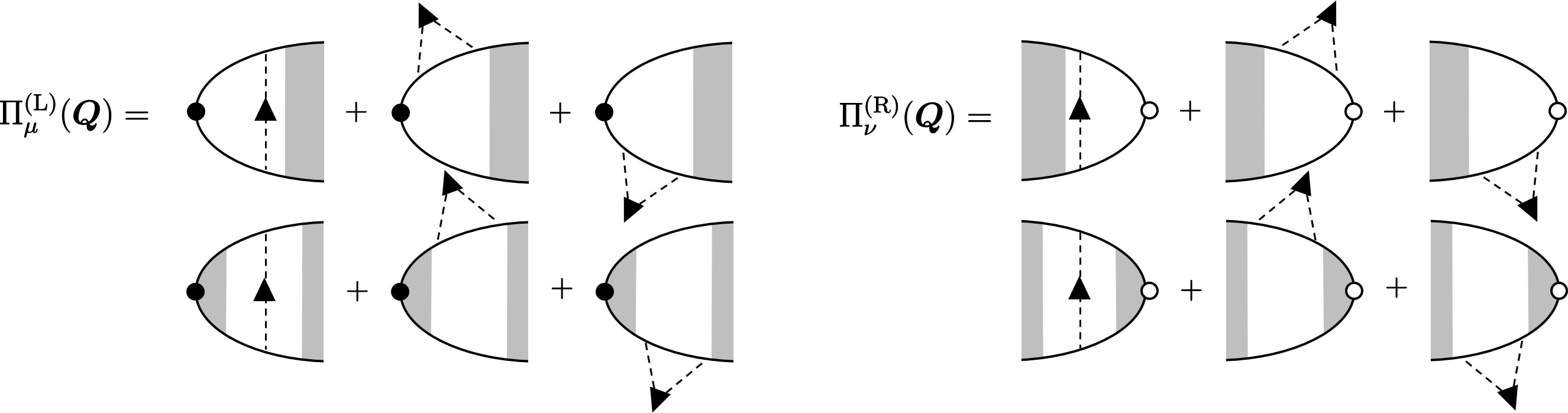}
  \caption{Left and right three-point vertices in the first order of $\delta n_{\rm i}$. 
   Arrows on the electron lines are suppressed for simplicity.}
  \label{3vc_lr}
 \end{center}
\end{figure}

\section{Response functions}

 In this Appendix, the response functions are calculated with the help of the following integrals, 
\begin{align}
 & I_{\mu \nu}(\bm q) \equiv \sum_{\bm k} v_{\mu}v_{\nu} G^{\text R}_{\bm k+}  G^{\text A}_{\bm k-}  ,
\label{app:I}
\\ 
 &Y_{\mu \nu}(\bm q, \bm Q) 
\equiv i\pi N(\mu ) \sum_{\bm k} v_iv_j G^R_{\bm k++} G^A_{\bm k--} ( G^R_{\bm k+-} - G^A_{\bm k-+} ) , 
\label{app:Y}
\\
& X_{\mu j}^{\text{sj(a)}, \alpha}(\bm q ) 
=e\frac{i\omega}{2\pi} \alpha^{\text{sj}}_{\text{SH}} \epsilon_{\alpha jl} I_{\mu l}(\bm q), 
\label{app:Xsj_a}
\\
& X_{i\nu}^{\text{sj(b)},\alpha}(\bm q ) 
=-e \frac{i\omega}{2\pi} \alpha^{\text{sj}}_{\text{SH}} \epsilon_{\alpha il} I_{\nu l}(\bm q), 
\label{app:Xsj_b}
\\
& X^{\text{sj}(c),\alpha}_{\mu \nu}(\bm q) 
= -e\frac{i\omega}{2\pi}  \alpha_{\text{SH}}^{\text{sj}} 
      \frac{1}{2\pi N(\mu )}  \epsilon_{\alpha lm} iq_l
     \Bigl[ I_{\mu m}(\bm q) I_{\nu}(\bm q) - I_{\mu} (\bm q) I_{\nu m}(\bm q) \Bigr] , 
\label{app:Xsj_c}
\\
& X^{\text{ss},\alpha}_{\mu \nu} (\bm q) 
= i\omega \alpha^{\text{ss}}_{\text{SH}} \frac{\sigma_{\text c}}{-e} 
    \left(  \frac{3 \gamma }{v_{\text F}^2 \pi N(\mu )} \right)^2 
    \epsilon_{\alpha lm} I_{\mu l} (\bm q) I_{\nu m} (\bm q) , 
\label{app:Xss}
\end{align}
where $v_\mu = 1$ for $\mu=0$ and $=v_i$ for $\mu = i = x,y,z$. 
 The Green's functions are denoted by 
$G^{\text{R/A}}_{\bm k ss'}=G^{\text{R/A}}_{\bm k + s\frac{\bm q}{2}+s'\frac{\bm Q}{2}}(s\frac{\omega}{2})$ 
 and $G^{\text{R/A}}_{\bm k s}=G^{\text{R/A}}_{\bm k+s\frac{\bm q}{2}}(s \frac{\omega}{2})$ 
with $s, s'=\pm $. 
 The integrals $X_{\mu \nu}^{\rm sj}$ will be used for the side-jump type processes, 
and $X_{\mu \nu}^{\rm ss}$ for the skew-scattering type processes. 
 The results of the integration are given in Appendix \ref{int}.

\subsection{Side-jump process\label{app_side-jump}}

The response functions of side-jump process contributed by the inhomogeneous scattering due to impurities (see Fig.~\ref{wovc_l_sj}) are derived as
\begin{align}
L^{\text{sj(c)},\alpha}_{\mu \nu} &= -e\frac{i\omega}{2\pi } \alpha^{\text{sj}}_{\text{SH}} \frac{\delta n_{\text i}(\bm Q)}{n_{\text i}} \frac{\gamma}{\pi N(\mu )}   
\epsilon_{\alpha lm}\tau \Bigl[ iq_l I_{\mu m}(\bm q+\bm Q)I_{\nu}(\bm q) - i(q_l+Q_l) I_{\mu }(\bm q+\bm Q) I_{\nu m}(\bm q) \Bigr] ,
\\
L^{\text{sj(d)},\alpha}_{\mu j} &= -e \frac{i\omega}{2\pi} \alpha^{\text{sj}}_{\text{SH}} \frac{\delta n_{\text i}(\bm Q)}{n_{\text i}} \frac{\gamma}{\pi N(\mu )} \epsilon_{\alpha lj}
\Bigl[  I_{\mu} (\bm q+\bm Q) I_{l} (\bm q) - Y_{\mu l} (\bm Q) \Bigr] ,
\\
L^{\text{sj(e)},\alpha}_{i \nu} &= -e \frac{i\omega}{2\pi} \alpha^{\text{sj}}_{\text{SH}} \frac{\delta n_{\text i}(\bm Q)}{n_{\text i}} \frac{\gamma}{\pi N(\mu )} \epsilon_{\alpha il}
\Bigl[ I_l (\bm q+\bm Q) I_{\nu} (\bm q) - Y_{l\nu} (\bm Q) \Bigr] ,
\\
L^{\text{sj(f)},\alpha}_{\mu \nu} &= -e\frac{i\omega}{2\pi} \alpha^{\text{sj}}_{\text{SH}} \frac{\delta n_{\text i}(\bm Q)}{n_{\text i}} \left( \frac{\gamma}{\pi N(\mu )} \right)^2 \epsilon_{\alpha lm} \tau
\biggl[
  iq_l \biggl\{
\Bigl(   I_{\nu } I_m - I_{\nu m} I_0 \Bigr)_{\bm q}
I_{\mu}(\bm q+\bm Q)
- \Bigl( Y_{\mu m}(\bm Q) I_{\nu}(\bm q) - Y_{\mu}(\bm Q) I_{\nu m}(\bm q) \Bigr) 
\biggr\}
\nonumber
\\
&
+
i(q_l+Q_l)
\bigg\{
 \Bigl(  I_{\mu m} I_0 - I_{\mu } I_m \Bigr)_{\bm q+\bm Q} 
I_{\nu}(\bm q)  
-\Bigl( I_{\mu m}(\bm q+\bm Q)Y_{\nu} - I_{\mu}(\bm q+\bm Q) Y_{\nu m} \Bigr) 
\biggr\}
\biggr] ,
\end{align}

\begin{figure}[tbp]
 \begin{center}
  \includegraphics[width=140mm]{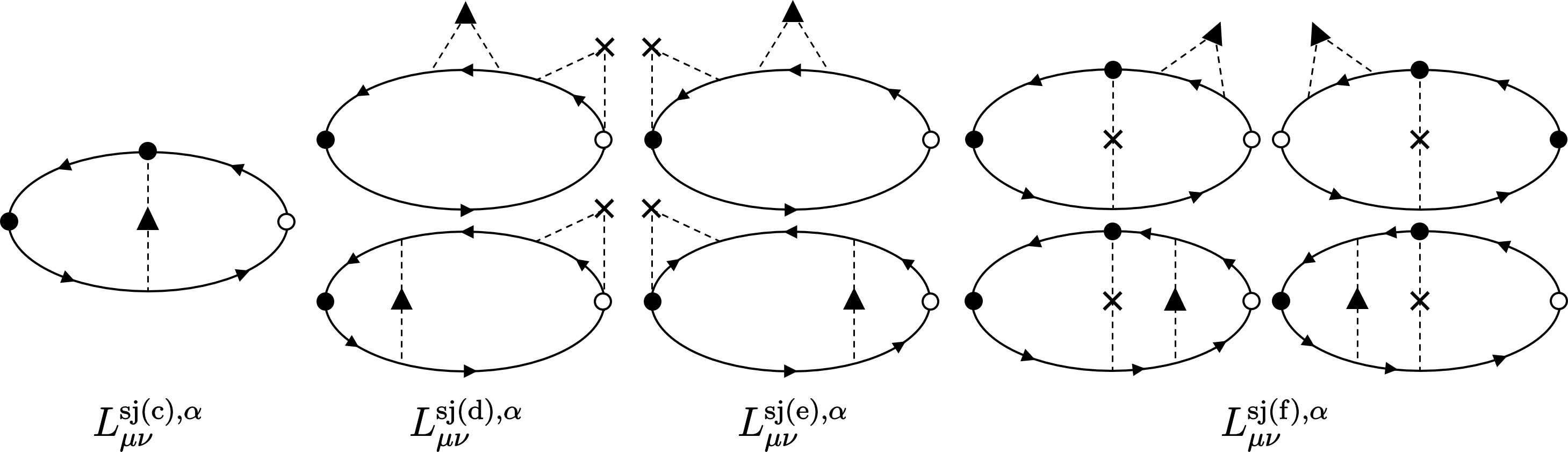}
  \caption{  
 Definition of $L^{\text{sj}, \alpha}_{\mu \nu}$ that contributes to the side-jump process  
shown in Fig.~\ref{sj_mod}.  
 The self-energy type and side-jump type parts can also be located at 
the lower line of the diagram. 
 All such possible patterns are considered in the calculation.}
  \label{wovc_l_sj}
 \end{center}
\end{figure}

With the ladder vertex corrections included, the response functions are expressed as
\begin{align}
K^{\text{sj(a-c)}, \alpha}_{0 j} &=\Bigl[ 1 + \Lambda^{\text s}_0(\bm q+\bm Q) \Bigr] \left[ L^{\text{sj(c)},\alpha}_{0j} + L^{\text{sj(c)},\alpha}_{00}\Lambda^{\text c}_j(\bm q) + \frac{\delta n_{\text i}(\bm Q)}{n_{\text i}} X^{\text{sj(a)},\alpha}_{0j} (\bm q+\bm Q) \right]  ,
\\
K^{\text{sj(a-c)}, \alpha}_{i j} &= \frac{\delta n_{\text i}(\bm Q)}{n_{\text i}}\Bigl[ X^{\text{sj(a)},\alpha}_{ij} (\bm q+\bm Q) + X^{\text{sj(b)},\alpha}_{ij} (\bm q) \Bigr]  
+ \left[ L^{\text{sj(c)},\alpha}_{i0} + \frac{\delta n_{\text i}(\bm Q)}{n_{\text i}} X^{\text{sj(b)},\alpha}_{i0} (\bm q) \right] \Lambda^{\text c}_j (\bm q)
\nonumber
\\
&
+ \Lambda^{\text s}_i(\bm q+\bm Q)  \left[ L^{\text{sj(c)},\alpha}_{0j} + \frac{\delta n_{\text i}(\bm Q)}{n_{\text i}} X^{\text{sj(a)},\alpha}_{0j} (\bm q+\bm Q) \right] , 
\\
K^{\text{sj(d-f)},\alpha}_{0j} &=\Bigl[ 1 + \Lambda^{\text s}_0(\bm q+\bm Q) \Bigr] \Bigl( L^{\text{sj(d)}, \alpha}_{0 j} + L^{\text{sj(f)}, \alpha}_{0 j} + L^{\text{sj(f)}, \alpha}_{0 0} \Lambda^{\text c}_j(\bm q) \Bigr)  + \Pi^{\text{(L)}}_0(\bm Q)  \Bigl[ X^{\text{sj(a)}, \alpha}_{0 j} (\bm q) + X^{\text{sj(c)}, \alpha}_{0 j} (\bm q) \Bigr] ,
\\
K^{\text{sj(d-f)},\alpha}_{ij} &= L^{\text{sj(d)}, \alpha}_{i j}+L^{\text{sj(e)}, \alpha}_{i j} 
+\Bigl( L^{\text{sj(e)}, \alpha}_{i0} +L^{\text{sj(f)}, \alpha}_{i0} \Bigr) \Lambda^{\text c}_j(\bm q) + \Bigl[ X^{\text{sj(b)}, \alpha}_{i0}(\bm q+\bm Q) + X^{\text{sj(c)}, \alpha}_{i0}(\bm q+\bm Q)  \Bigr]\Pi^{\text{(R)}}_j(\bm Q)
\nonumber
\\
&+ \Lambda^{\text s}_i(\bm q+\bm Q) \Bigl( L^{\text{sj(d)}, \alpha}_{0 j} + L^{\text{sj(f)}, \alpha}_{0 j} \Bigr) + \Pi^{\text{(L)}}_i(\bm Q)  \Bigl[ X^{\text{sj(a)}, \alpha}_{0 j} (\bm q) + X^{\text{sj(c)}, \alpha}_{0 j} (\bm q) \Bigr] 
.
\end{align}
The physical quantities are obtained as
\begin{align}
\average{\sigma^{\alpha}}^{\text{sj(a-c)}}_{\delta} 
&= - \alpha^{\text{sj}}_{\text{SH}} \frac{\delta n_{\text i}(\bm Q)}{n_{\text i}} \frac{\sigma_{\text c}}{-e} \frac{[i(\bm q+\bm Q) \times \bm E]_{\alpha}}{D(\bm q+\bm Q)^2 -i\omega +\tau_{\text{sf}}^{-1}} 
 + \alpha^{\text{sj}}_{\text{SH}} \frac{\delta n_{\text i}(\bm Q)}{n_{\text i}}\frac{\sigma_{\text c}}{-e} \frac{\epsilon_{\alpha lm}iQ_l}{D(\bm q+\bm Q)^2-i\omega + \tau_{\text{sf}}^{-1}} 
    \frac{D q_m (\bm q\cdot \bm E)}{Dq^2-i\omega} ,
 \label{app_sj_spin_abc}
\\
\average{j_{\text s,i}^{\alpha}}^{\text{sj(a-c)}}_{\delta} 
&= \alpha^{\text{sj}}_{\text{SH}} \frac{\delta n_{\text i}(\bm Q)}{n_{\text i}} \epsilon_{\alpha ij} \frac{\average{j_{\text e, j}(\bm q)}_0}{-e} - Di(q_i+Q_i) \average{\sigma^{\alpha}}_{\delta}^{\text{sj(a-c)}},
\label{app_sj_sc_abc}
\\
\average{\sigma^{\alpha}}^{\text{sj(d-f)}}_{\delta} 
&= \alpha^{\text{sj}}_{\text{SH}} \frac{\delta n_{\text i} (\bm Q)}{n_{\text i}} \frac{\sigma_{\text c}}{-e}  \frac{1}{D(\bm q+\bm Q)^2-i\omega + \tau_{\text{sf}}^{-1}}
\left[ [i(\bm q+\bm Q)\times \bm E]_{\alpha} - \frac{D\bm q\cdot (\bm q+\bm Q) [i\bm q\times \bm E]_{\alpha}}{Dq^2-i\omega + \tau_{\text{sf}}^{-1}} \right] 
\nonumber
\\
 &- \alpha^{\text{sj}}_{\text{SH}} \frac{\delta n_{\text i}(\bm Q)}{n_{\text i}}\frac{\sigma_{\text c}}{-e} \frac{\epsilon_{\alpha lm}iQ_l}{D(\bm q+\bm Q)^2-i\omega + \tau_{\text{sf}}^{-1}} 
 \frac{D \, q_m (\bm q \!\cdot\! \bm E)}{Dq^2-i\omega},
 \label{app_sj_spin_def}
\\
\average{j_{\text s,i}^{\alpha}}^{\text{sj(d-f)}}_{\delta} 
&=  \alpha^{\text{sj}}_{\text{SH}} \epsilon_{\alpha ij} \frac{\average{j_{\text e, j}(\bm q+\bm Q)}_{\delta}}{-e} 
-Di(q_i+Q_i) \average{\sigma^{\alpha}}^{\text{sj(d-f)}}_{\delta} + Diq_i \frac{\delta n_{\text i}(\bm Q)}{n_{\text i}} \average{\sigma^{\alpha}}^{\text{sj}}_0.
\label{app_sj_sc_def}
\end{align}
 Equations (\ref{app_sj_spin_abc}) and (\ref{app_sj_sc_abc}) are due to the inhomogeneity 
of SOI  scattering , 
and Eqs.~(\ref{app_sj_spin_def}) and (\ref{app_sj_sc_def}) are due to the inhomogeneity 
of normal  scattering .

\subsection{Skew scattering process\label{app_skew_scattering}}

 The response function $L^{\text{ss}, \alpha}_{\mu \nu}$ due to skew scattering 
from the inhomogeneous impurities, shown in Fig.~\ref{wovc_l_ss}, is expressed as
\begin{align}
 L^{\text{ss}, \alpha}_{\mu \nu} =
i\omega \alpha^{\text{ss}}_{\text{SH}} \frac{\delta n_{\text i}(\bm Q)}{n_{\text i}}  \frac{\sigma_{\text c}}{-e}   \left( \frac{3}{v_{\text F}^2} \right)^2 \left( \frac{\gamma}{\pi N(\mu )} \right)^3   \epsilon_{\alpha lm} 
 \biggl\{
 \Bigl[ 
&I_{\mu}(\bm q+\bm Q) I_l(\bm q) - Y_{\mu l} (\bm Q) \Bigr]
I_{\nu m} (\bm q) 
\nonumber
\\
+&I_{\mu l}(\bm q+\bm Q)
\Bigl[ I_m(\bm q+\bm Q) I_{\nu }(\bm q) - Y_{\nu m}(\bm Q) \Bigr] 
\biggr\}.
\end{align}

\begin{figure}[tbp]
 \begin{center}
  \includegraphics[width=140mm]{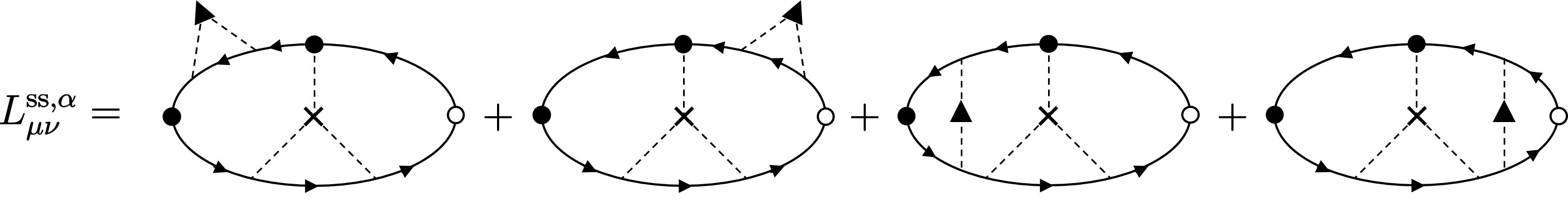}
  \caption{ 
 Definition of $L^{\text{ss}, \alpha}_{\mu \nu}$ that contributes to the skew-scattering 
process shown in Fig.~\ref{skew_mod}. 
 As in the case of Fig.~\ref{wovc_l_sj}, all possible patterns as to the position of the self-energy 
type part are considered in the calculation.
}
  \label{wovc_l_ss}
 \end{center}
\end{figure}

 With the ladder vertex corrections included, the response functions are given as
\begin{align}
 K^{\text{ss(a)}, \alpha}_{0j} 
&= i\omega \alpha^{\text{ss}}_{\text{SH}} \frac{\delta n_{\text i}(\bm Q)}{n_{\text i}} \frac{\sigma_{\text c}}{-e}  \left( \frac{3\gamma}{v_{\text F}^2\pi N(\mu )} \right)^2
 \epsilon_{\alpha lm}\Bigl[ 1+ \Lambda_{0}^{\text s}(\bm q+\bm Q)\Bigr] I_{l} (\bm q+\bm Q) \Bigl[ I_{ mj} (\bm q)+I_m(\bm q) \Lambda^{\text c}_{j}(\bm q)  \Bigr],
\\
 K^{\text{ss(a)}, \alpha}_{ij} 
&= i\omega \alpha^{\text{ss}}_{\text{SH}} \frac{\delta n_{\text i}(\bm Q)}{n_{\text i}} \frac{\sigma_{\text c}}{-e}  \left( \frac{3\gamma}{v_{\text F}^2\pi N(\mu )} \right)^2
 \epsilon_{\alpha lm} 
 \Bigl[ I_{il} (\bm q+\bm Q) +\Lambda_{i}^{\text s}(\bm q+\bm Q)I_{l} (\bm q+\bm Q)  \Bigr] 
 \Bigl[ I_{ mj} (\bm q)+I_{ m} (\bm q) \Lambda_{j}^{\text c}(\bm q)  \Bigr] 
,
\\
 K^{\text{ss(b)}, \alpha}_{0j} 
&= \Bigl[ 1+\Lambda^{\text s}_0(\bm q+\bm Q) \Bigr]  \Bigl[ L^{\text{ss}, \alpha}_{0j}+L^{\text{ss}, \alpha}_{00}\Lambda^{\text c}_j(\bm q)\Bigr]
+ \Pi^{\text{(L)}}_0(\bm Q) X^{\text{ss}, \alpha}_{0j}(\bm q),
\\
 K^{\text{ss(b)}, \alpha}_{ij} 
&= L^{\text{ss}, \alpha}_{ij} + L^{\text{ss}, \alpha}_{i0} \Lambda^{\text c}_j(\bm q) + X^{\text{ss}}_{i0}(\bm q +\bm Q) \Pi^{\text{(R)}}_j(\bm Q) 
+ \Lambda^{\text s}_i (\bm q+\bm Q) \Bigl[ L^{\text{ss}, \alpha}_{0j}+L^{\text{ss}, \alpha}_{00}\Lambda^{\text c}_j(\bm q)\Bigr] + \Pi^{\text{(L)}}_i(\bm Q) X^{\text{ss},\alpha}_{0j}(\bm q).
\end{align}
The spin accumulation and spin current coming from each contributions are derived as
\begin{align}
\average{\sigma^{\alpha}}_{\delta}^{\text{ss(a)}} &= -\alpha^{\text{ss}}_{\text{SH}} \frac{\delta n_{\text i}(\bm Q)}{n_{\text i}} \frac{\sigma_{\text c}}{-e} \frac{[i(\bm q+\bm Q) \times \bm E]_{\alpha}}{D(\bm q+\bm Q)^2 -i\omega +\tau_{\text{sf}}^{-1}}
 + \alpha^{\text{ss}}_{\text{SH}} \frac{\delta n_{\text i}(\bm Q)}{n_{\text i}}\frac{\sigma_{\text c}}{-e} \frac{\epsilon_{\alpha lm}iQ_l}{D(\bm q+\bm Q)^2-i\omega + \tau_{\text{sf}}^{-1}} 
   \frac{Dq_m (\bm q\cdot \bm E)}{Dq^2-i\omega},
 \label{app_skew_spin_a}
\\
\average{j_{\text s, i}^{\alpha}}_{\delta}^{\text{ss(a)}} &= \alpha^{\text{ss}}_{\text{SH}} \frac{\delta n_{\text i}(\bm Q)}{n_{\text i}} \frac{\sigma_{\text c}}{-e} \epsilon_{\alpha ij} \frac{\average{j_{\text e, j}(\bm q)}_0}{-e} - Di(q_i+Q_i) \average{\sigma^{\alpha}}_{\delta}^{\text{ss(a)}},
\label{app_skew_sc_a}
\\
\average{\sigma^{\alpha}}_{\delta}^{\text{ss(b)}} &= \alpha^{\text{ss}}_{\text{SH}} \frac{\delta n_{\text i}(\bm Q)}{n_{\text i}} \frac{\sigma_{\text c}}{-e}  \frac{1}{D(\bm q+\bm Q)^2 -i\omega +\tau_{\text{sf}}^{-1}} 
\Biggl[
2[i(\bm q+\bm Q) \times \bm E]_{\alpha} - \frac{D\bm q\cdot (\bm q+\bm Q) [ i\bm q\times \bm E]_{\alpha}}{Dq^2-i\omega + \tau_{\text{sf}}^{-1}}
\Biggr] 
\nonumber
\\
&-  2\alpha^{\text{ss}}_{\text{SH}} \frac{\delta n_{\text i}(\bm Q)}{n_{\text i}}\frac{\sigma_{\text c}}{-e} \frac{\epsilon_{\alpha lm}iQ_l}{D(\bm q+\bm Q)^2-i\omega + \tau_{\text{sf}}^{-1}} 
   \frac{D \, q_m (\bm q \!\cdot\! \bm E) }{Dq^2-i\omega},
\label{app_skew_spin_b}
\\
\average{j^{\alpha}_{\text s, i}}_{\delta}^{\text{ss(b)}} &= -\alpha^{\text{ss}}_{\text{SH}} \frac{\delta n_{\text i}(\bm Q)}{n_{\text i}} \epsilon_{\alpha ij} \frac{\average{j_{\text e, j}(\bm q)}_0}{-e} + \alpha^{\text{ss}}_{\text{SH}} \epsilon_{\alpha ij} \frac{\average{j_{\text e, j}(\bm q+\bm Q)}_{\delta}}{-e} 
-Di(q_i+Q_i) \average{\sigma^{\alpha}}^{\text{ss(b)}}_{\delta} + Diq_i \frac{\delta n_{\text i}(\bm Q)}{n_{\text i}} \average{\sigma^{\alpha}(\bm q)}^{\text{ss}}_0.
\label{app_skew_sc_b}
\end{align}
 Equations (\ref{app_skew_spin_a}) and (\ref{app_skew_sc_a}) are due to the inhomogeneity of SOI impurities, 
and Eqs.~(\ref{app_skew_spin_b}) and (\ref{app_skew_sc_b}) are due to the inhomogeneity of normal impurities.
\end{widetext}

\subsection{Extrinsic Rashba process}

The response functions of the ``extrinsic Rashba process" (Sec.~\ref{ex-rashba}) are derived from the diagrams shown in Fig.~\ref{1st}, 
\begin{align}
K^{\text{R(a)},\alpha}_{\mu j} &=  e\lambda_{\text{so}} \delta n_{\text i}(\bm Q) u_{\text i} \frac{i\omega }{\pi}  \epsilon_{\alpha lj} iQ_l 
\sum_{\bm k} v_{\mu} 
G^{\text R}_{\bm k++} G^{\text A}_{\bm k--},
\\
K^{\text{R(b)},\alpha}_{i\nu} &=  -e\lambda_{\text{so}} \delta n_{\text i}(\bm Q) u_{\text i} \frac{i\omega }{\pi}  \epsilon_{\alpha il} iQ_l \sum_{\bm k} v_{\nu} 
G^{\text R}_{\bm k++} G^{\text A}_{\bm k--},
\\
K^{\text{R(c)},\alpha}_{\mu \nu} &=  e\lambda_{\text{so}} \delta n_{\text i} (\bm Q) u_{\text i} \frac{i\omega }{\pi} \epsilon_{\alpha lm} iQ_l 
  \sum_{\bm k} v_{\mu}v_{\nu} 
G^{\text R}_{\bm k++} G^{\text A}_{\bm k--}
\nonumber
\\ & \times
\left[ \left( k+\frac{q}{2} \right)_m G^{\text R}_{\bm k+-} 
+ \left( k-\frac{q}{2} \right)_{m} G^{\text A}_{\bm k-+} \right] .
\end{align}
With the ladder vertex corrections included, their sum turns out to vanish.

\begin{figure}[tb]
 \begin{center}
  \includegraphics[width=80mm]{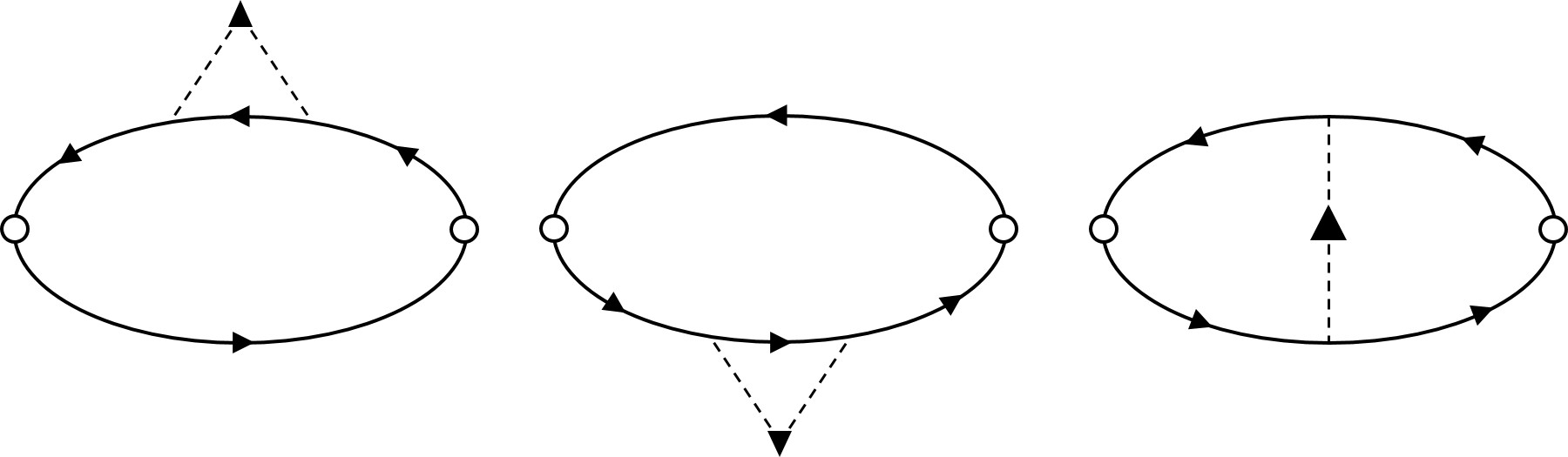}
  \caption{Contributions to the response function $L^{\text{cc}, \alpha}_{\mu \nu}$.}
  \label{fig_cc}
 \end{center}
\end{figure}

\begin{figure}[tb]
 \begin{center}
  \includegraphics[clip,width=80mm]{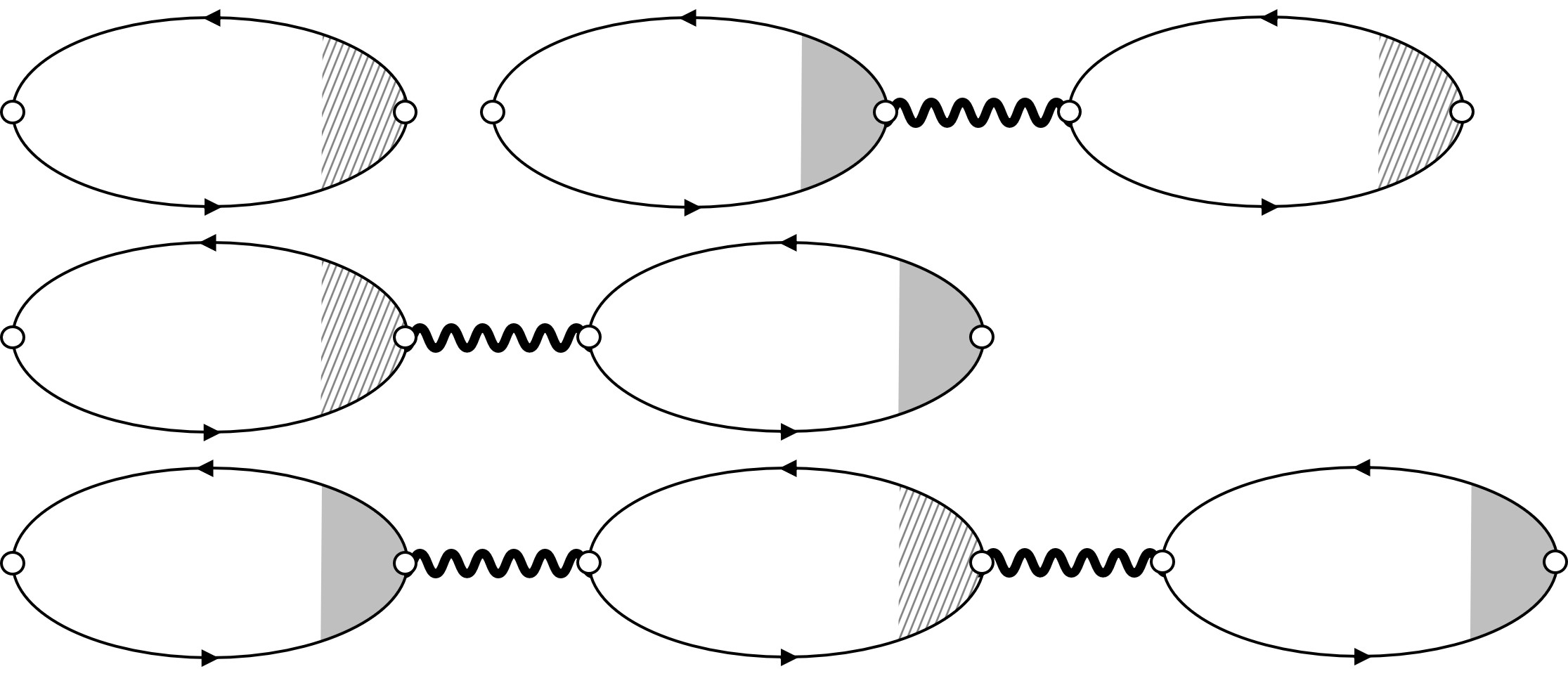}
  \caption{
Charge and current densities in the presence of long-range Coulomb interaction. 
  The bold wavy line represents the effective Coulomb interaction 
defined by Fig.~\ref{v_eff} or Eq.~(\ref{v_eff_rpa}).
  The diagrams without ladder vertex corrections are also included in the calculation.}
  \label{rpa_e}
 \end{center}
\end{figure}

\subsection{Charge density and current density}

We define the response functions of charge and current densities as 
\begin{align}
\average{j_{\text e, \mu}(\bm q+\bm Q)}_{\delta}=K^{\text{cc}}_{\mu \nu} A_{\nu}(\bm q, \omega).
\end{align} 
The response functions of charge and current densities without ladder vertex corrections 
(see Fig.~\ref{fig_cc}) are derived as
\begin{align}
L^{\text{cc}}_{\mu \nu} = e^2\frac{i\omega}{\pi} \frac{\delta n_{\text i}(\bm Q)}{n_{\text i}} n_{\text i}u_{\text i}^2 
\Bigl[   I_{\mu }(\bm q+\bm Q) I_{\nu}(\bm q)-Y_{\mu \nu}(\bm Q) \Bigr] . 
\end{align} 
 With the ladder vertex corrections included, the response functions are derived as
\begin{align}
 K^{\text{cc}}_{0j} 
&= 
  L_{0j}^{\text{cc}} + L_{00}^{\text{cc}} \Lambda _j^{\text c} (\bm q)  
+ e^2\frac{i\omega}{\pi}  I_0(\bm q+\bm Q) \Pi^{\text{(R)}}_j (\bm Q) ,
\\
 K^{\text{cc}}_{ij} &=L_{ij}^{\text{cc}}  +
  L_{i0}^{\text{cc}} \Lambda^{\text c}_j (\bm q) 
 + e^2\frac{i\omega}{\pi}  I_i(\bm q+\bm Q) \Pi^{\text{(R)}}_j (\bm Q) .
\end{align}

\subsection{Results}
\label{App:results}

 The results are summarized as follows, 
\begin{align}
 \langle \hat {\rho}_{\rm e} \rangle_0 
&= -\sigma_{\rm c} \Pi_{\rm c} ({\bm q}) (i{\bm q}\cdot {\bm E})   ,
\label{cd_0_result}
\\
 \langle \hat {\bm j}_{\text e} \rangle_0 
&= \sigma_{\rm c} {\bm F}  , 
\label{cc_0_result}
\\
  \average{ \hat \sigma^\alpha}_0  
&= - \frac{\sigma_{\rm c}}{-e}  \Pi_{\rm s}({\bm q}) \alpha_{\rm SH} 
    [ i{\bm q} \times {\bm E} ]_\alpha  , 
\label{sd_0_result}
\\
  \average{ \hat j^\alpha_{\text s,i}}_0  
&= \frac{\sigma_{\rm c} }{-e} \alpha_{\rm SH}  \left\{  \epsilon_{\alpha ij} F_j  
  - D q_i  \Pi_{\rm s}({\bm q}) [ {\bm q} \times {\bm E} ]_\alpha  \right\} , 
\label{sc_0_result}
\end{align}

\begin{widetext}

\begin{align}
 \average{ \hat \rho_{\rm e}}_\delta 
&= - \delta \sigma_{\rm c}  
    \Pi_{\rm c} ({\bm q}+{\bm Q}) [ i({\bm q} + {\bm Q}) \!\cdot\! {\bm F} ]   ,
\label{cd_result}
\\
 \average{\hat {\bm j_{\rm e}}}_\delta 
&=  \delta \sigma_{\rm c} 
    \left\{ {\bm F} 
 - [ 1 + i\omega   \Pi_{\rm c} ({\bm q}+{\bm Q}) ] 
   \frac{({\bm q} + {\bm Q})[ ({\bm q}+{\bm Q}) \!\cdot\! {\bm F} ] }{ ({\bm q}+{\bm Q})^2}   
   \right\}  , 
\label{cc_result}
\\
  \average{\hat \sigma^\alpha}_\delta  
&= \frac{\delta \sigma_{\rm c}}{-e}  
    \Pi_{\rm s}({\bm q}+{\bm Q}) \biggl\{ 
   - \alpha^{\rm ss}_{\rm SH} [ i ({\bm q} +{\bm Q}) \times  {\bm F}   ]_\alpha   
 + \alpha_{\rm SH}  D \, [ ({\bm q}+{\bm Q}) \!\cdot\! {\bm q}] \, 
   \Pi_{\rm s}({\bm q})  [ i{\bm q} \times {\bm F} ]_\alpha 
   \biggr\} , 
\label{sd_result}
\\ 
  \average{\hat j^\alpha_{\text s,i}}_\delta  
&= \frac{\delta \sigma_{\rm c}}{-e} \alpha_{\rm SH}^{\rm ss} 
    \Bigl\{  \epsilon_{\alpha ij}  F_j  
   - D (q_i+Q_i) \Pi_{\rm s}({\bm q}+{\bm Q}) [ ({\bm q} + {\bm Q}) \times {\bm F}]^\alpha  \Bigr\}   
\nonumber \\
& +  \frac{\delta \sigma_{\rm c}}{-e} \alpha_{\rm SH}  \Bigl\{  
  - \epsilon_{\alpha ij} \, [1 + i\omega \Pi_{\rm c} ({\bm q}+{\bm Q})] \frac{q_j+Q_j}{({\bm q} + {\bm Q})^2}
    [({\bm q} + {\bm Q}) \!\cdot\! {\bm F} ]   
\nonumber \\
&\hskip 18mm 
   -  D \bigl[ q_i -  (q_i+Q_i)   
    \Pi_{\rm s}({\bm q}+{\bm Q})   D \, [ ({\bm q}+{\bm Q}) \!\cdot\! {\bm q}] \,  \bigr] \Pi_{\rm s}({\bm q}) ( {\bm q} \times {\bm E} )_\alpha 
   \Bigr\}  . 
\label{sc_result}
\end{align}
\end{widetext}
 Here, $\Pi_{\rm s}({\bm q}) = (Dq^2 - i\omega + \tau_{\text{sf}}^{-1})^{-1}$ 
is the spin diffusion propagator,  
$\Pi_{\rm c}({\bm q})$ is the charge diffusion propagator (see below), 
and $\delta \sigma_{\rm c} = - (\delta n_{\rm i}/n_{\rm i}) \sigma_{\rm c} $ 
is the modulation of electrical conductivity.  
 We defined the effective field, 
\begin{align}
 {\bm F} &\equiv  {\bm E}  - [ 1 + i\omega  \Pi_{\rm c} ({\bm q}) ] \,  {\bm E}_\parallel 
\label{F_def1}
\\
&= {\bm E}_\perp  - i\omega  \Pi_{\rm c} ({\bm q}) \,  {\bm E}_\parallel  , 
\label{F_def2}
\end{align}
that induces the charge current, Eq.~(\ref{cc_0_result}) (which defines ${\bm F}$).
 While the first term of Eq.~(\ref{F_def1}) is the real (applied) electric field, 
the second term just expresses the effects of charge accumulation 
produced by the longitudinal component ${\bm E}_\parallel$ 
[see Eq.~(\ref{F_wo_C}) or Eq.~(\ref{F_Coulomb})]. 
 Note that ${\bm q} \times {\bm F} = {\bm q} \times {\bm E}$. 
 Let us express Eq.~(\ref{F_def1}) as 
\begin{align}
  {\bm F} &= {\cal P} ({\bm q}) {\bm E} , 
\end{align}
with a matrix ${\cal P} ({\bm q})$, 
\begin{align}
 {\cal P}_{ij} ({\bm q}) &= \delta_{ij} - [ 1 + i\omega  \Pi_{\rm c} ({\bm q}) ] \, \frac{q_iq_j}{q^2}  . 
\label{eq:P}
\end{align}
 Then, Eq.~(\ref{cc_result}) is expressed as 
\begin{align}
 \average{{\bm j_{\rm e}}}_\delta 
&= \delta \sigma_{\rm c} {\cal P} ({\bm q}+{\bm Q}) {\bm F} 
\nonumber \\
&= \delta \sigma_{\rm c} {\cal P} ({\bm q}+{\bm Q}) {\cal P} ({\bm q}) {\bm E}  . 
\label{cc_PP}
\end{align}
 Combined with Eq.~(\ref{cc_0_result}), this may be written as 
\begin{align}
  \delta (\sigma_{\rm c} {\bm F}) &= \delta \sigma_{\rm c} {\cal P} ({\bm q}+{\bm Q}) {\bm F} . 
\end{align}
where $\delta$ expresses the effect of impurity-concentration modulation. 
 Using this, we can \lq\lq derive'' the terms that have $\epsilon_{\alpha ij}$ in Eq.~(\ref{sc_result}). 
 From Eq.~(\ref{sc_0_result}), they are expected to be obtained as $\delta (\alpha_{\rm SH} \sigma_{\rm c} F_j) 
= \delta \alpha_{\rm SH}^{\rm sj} (\sigma_{\rm c} F_j) + \alpha_{\rm SH}  \delta (\sigma_{\rm c} F_j)
= - \alpha_{\rm SH}^{\rm sj} \delta \sigma_{\rm c} F_j + \alpha_{\rm SH}  \delta (\sigma_{\rm c} F_j)$,  
or
\begin{align}
& - \alpha_{\rm SH}^{\rm sj}  \, F_j  
   + \alpha_{\rm SH} [ {\cal P} ({\bm q}+{\bm Q}) {\bm F} \, ]_j    
\nonumber \\
&=  \alpha_{\rm SH}^{\rm ss}  F_j    
  -   \alpha_{\rm SH}   
   \, [1 + i\omega \Pi_{\rm c} ({\bm q}+{\bm Q})] \frac{q_j+Q_j}{({\bm q} + {\bm Q})^2}
    [({\bm q} + {\bm Q}) \!\cdot\! {\bm F} ] ,  
\end{align}
except for the overall factor of $\delta \sigma_{\rm c}$. 
 These are indeed the terms of interest.

 The statements so far presented in this section hold also for the case 
with long-range Coulomb interaction (given in Appendix \ref{app_coulomb}). 
 In the present case (without Coulomb interaction), we have 
$\Pi_{\rm c}({\bm q}) = (Dq^2 - i\omega)^{-1}$, and 
${\cal P}_{ij} ({\bm q}) = \delta_{ij} - D q_i q_j \Pi_{\rm c} ({\bm q})$, or 
\begin{align}
 {\bm F} &=  {\bm E}  - Dq^2 \Pi_{\rm c}({\bm q}) {\bm E}_\parallel  . 
\label{F_wo_C}
\end{align}
 The second term of this equation corresponds to a diffusion current. 
 The corresponding expression in the presence of Coulomb interaction 
will be given as Eq.~(\ref{F_Coulomb}).

\section{Inclusion of Coulomb interaction}
\label{app_coulomb}

\subsection{Calculation} 

 Treating the Coulomb interaction in the random phase approximation (RPA), 
the response functions which are zeroth order and first order in $\delta n_{\text i}$ are calculated from 
\begin{align}
 K^{\text{RPA}}_{\mu j} 
&= K^{\text{cc}}_{\mu j} + K^{\text{cc}}_{\mu 0} (-U^{\text{eff}}_{\bm q}) \pi^{\text{cc}}_{0j}(\bm q )
\nonumber
\\
& + \pi^{\text{cc}}_{\mu 0}(\bm q+\bm Q ) (-U_{\bm q+\bm Q}^{\text{eff}}) K^{\text{cc}}_{0j}
\nonumber
\\
&+ \pi^{\text{cc}}_{\mu 0}(\bm q+\bm Q ) (-U_{\bm q+\bm Q}^{\text{eff}}) 
  K^{\text{cc}}_{00} (-U_{\bm q}^{\text{eff}}) \pi^{\text{cc}}_{0j}(\bm q )  , 
\end{align}
where 
\begin{align}
\pi^{\text{cc}}_{\mu \nu}(\bm q ) &\equiv \frac{i}{e^2} \int^{\infty}_0 dt e^{i(\omega +i\delta ) t} \average{[\hat j_{\text e, \mu} (\bm q, t), \hat j_{\text e, \nu} (-\bm q) ]} , 
\label{app:pi^cc}
\end{align}
is the electromagnetic response function, 
and the effective Coulomb interaction $U^{\text{eff}}_{\bm Q}$ is now given by 
\begin{align}
 U^{\text{eff}}_{\bm q} (\omega) 
= \frac{Dq^2-i\omega}{D(q^2+q_{\text T}^2) -i\omega} U_{\bm q} . 
\label{v_eff_rpa}
\end{align}
$K^{\text{cc}}_{\mu 0}$ are the response functions of charge density and current to the scalar potential given in Eqs.~(\ref{app_cc_00}) and (\ref{app_cc_i0}).
The response functions are given by
\begin{widetext}
\begin{align}
 K^{\text{RPA}}_{0j} 
&=  i\omega \sigma_{\text c} \frac{\delta n_{\text i}}{n_{\text i}} 
   \frac{1}{D[(\bm q+\bm Q)^2 + q_{\text T}^2] -i\omega} 
   \left[ i(q_j+Q_j) - \frac{D\bm q \cdot (\bm q+\bm Q) iq_j}{D(q^2+q_{\text T}^2)-i\omega} 
     \left( 1+\frac{q_{\text T}^2}{q^2} \right)  \right] ,
\\
 K^{\text{RPA}}_{ij} 
&= -i\omega \sigma_{\text c} \frac{\delta n_{\text i}}{n_{\text i}} 
   \left[ \delta_{ij} 
   - \frac{D q_i q_j}{D(q^2+q_{\text T}^2)-i\omega} \left( 1+\frac{q_{\text T}^2}{q^2} \right)
   \right]
  - Di(q_i+Q_i) K^{\text{RPA}}_{0j}  . 
\end{align}
 Therefore, the electric charge and current densities at first order in $\delta n_{\rm i}$ are given by
\begin{align}
&\average{\hat \rho_{\text e}}^{\text{RPA}}_{\delta} 
=  \sigma_{\text c} \frac{\delta n_{\text i}}{n_{\text i}} 
   \frac{1}{D[(\bm q+\bm Q)^2 + q_{\text T}^2] -i\omega} 
   \biggl[ i(\bm q+\bm Q) \cdot \left( \bm E + \bm E^{\text{ind}}_0   \right) 
    - \frac{D \bm q \cdot (\bm q+\bm Q) i\bm q\cdot \bm E}{D(q^2+q_{\text T}^2)-i\omega} \biggr] ,
\\
&\average{\hat{\bm j_{\text e}}}^{\text{RPA}}_{\delta} 
=  \sigma_{\text c} 
 \left[ - \frac{\delta n_{\text i}}{n_{\text i}} 
         \left( \bm E+\bm E^{\text{ind}}_0  \right) +  \bm E^{\text{ind}}_\delta  \right] 
  -\delta D(\bm Q) i\bm q \average{\hat \rho_{\text e}}_0^{\text{RPA}}
  - D i(\bm q + \bm Q) \average{\hat \rho_{\text e}}^{\text{RPA}}_{\delta},
\end{align}
\end{widetext}
where we defined the electric fields, 
\begin{align}
 \bm E^{\text{ind}}_0  &= -i\bm q \frac{\average{\hat \rho_{\text e}}_0^{\text{RPA}}}{\epsilon_0 q^2},
\\
 \bm E^{\text{ind}}_\delta  
&= -i(\bm q + \bm Q) 
    \frac{\average{\hat \rho_{\text e}}^{\text{RPA}}_{\delta}}{\epsilon_0 (\bm q+\bm Q)^2} , 
\end{align}
due to the charge accumulation, $\average{\hat \rho_{\text e}}^{\text{RPA}}_0$ and 
$\average{\hat \rho_{\text e}}^{\text{RPA}}_{\delta}$, respectively. 
 Here, those at zeroth order in $\delta n_{\text i}$ (in the absence of the impurity inhomogeneity) 
are given by
\begin{align}
\average{\hat \rho_{\text e}}^{\text{RPA}}_0 &= -\sigma_{\text c} \frac{i\bm q\cdot \bm E}{D(q^2+q_{\text T}^2) -i\omega},
\\
\average{\hat{\bm j}_{\text e}}_0^{\text{RPA}} &=  \sigma_{\text c}\left( \bm E + \bm E^{\text{ind}}_0  \right) -Di\bm q \average{\hat \rho_{\text e}}^{\text{RPA}}_0 ,
\end{align}
For calculation, we derive the response functions to the scalar potential given by
\begin{align}
 K^{\text{cc}}_{00} 
&= L^{\text {cc}}_{00}  \Bigl[ 1+\Lambda^{\text c}_0 (\bm q ) \Bigr] 
 + e^2\frac{i\omega}{\pi}  I_0(\bm q+\bm Q) \Pi^{\text{(R)}}_0 (\bm Q) ,
\label{app_cc_00}
\\
 K^{\text{cc}}_{i0} 
&= L^{\text {cc}}_{i0} \Bigl[ 1+\Lambda^{\text c}_0(\bm q) \Bigr] 
 + e^2\frac{i\omega}{\pi}  I_i (\bm q+\bm Q) \Pi^{\text{(R)}}_0 (\bm Q) .
\label{app_cc_i0}
\end{align}

\begin{figure}[tb]
 \begin{center}
  \includegraphics[clip,width=75mm]{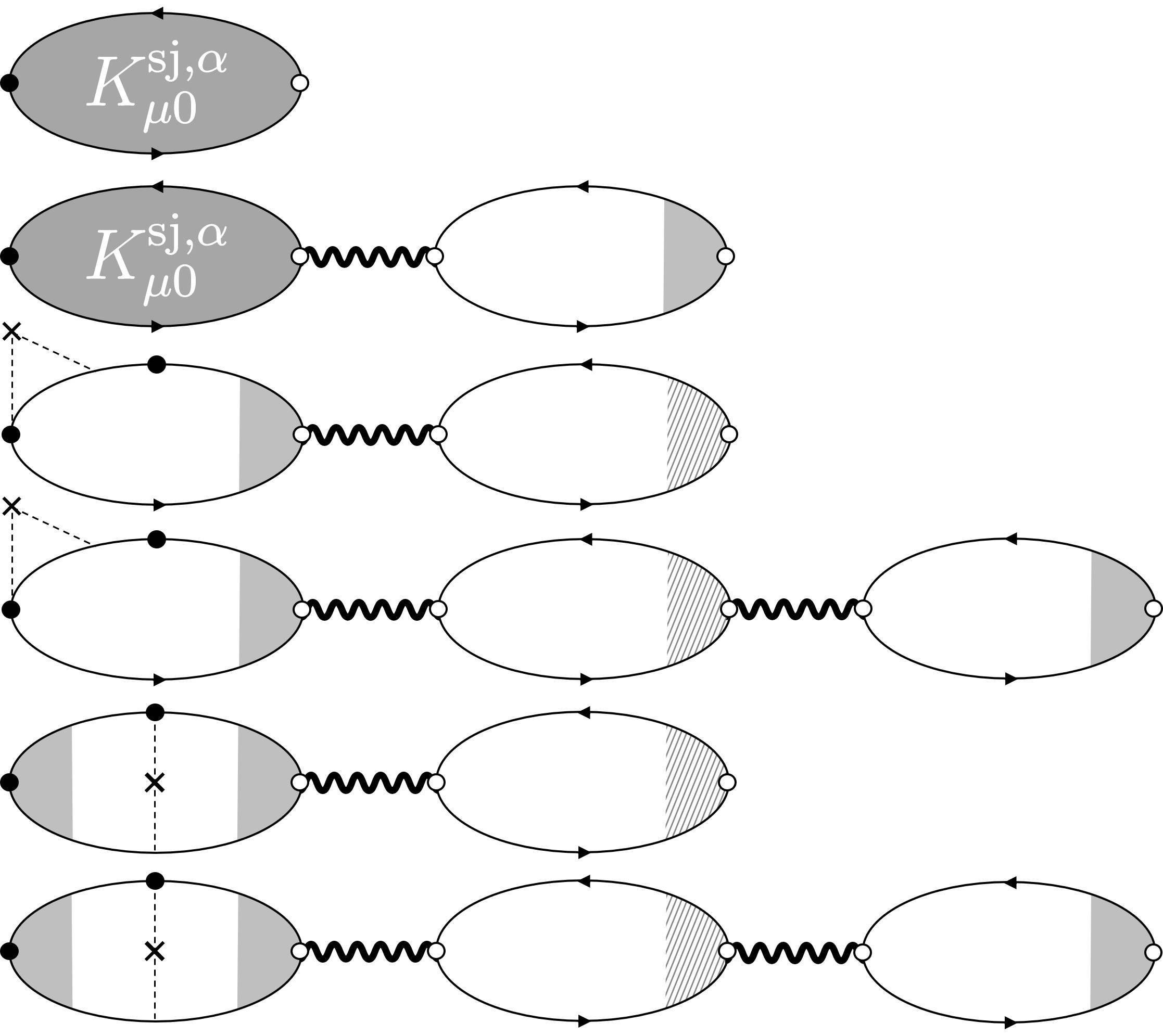}
  \caption{
 The side-jump contribution in the presence of long-range Coulomb interaction. 
  The leftmost gray ovals in the first and the second lines are the response functions 
in the absence of the Coulomb interaction, given by Eqs.~(\ref{app_sj_00}) and (\ref{app_sj_i0}), 
respectively [see Fig.~\ref{sj_mod}]. 
     The upside-down diagrams and the diagrams without ladder vertex corrections are also included in the calculation.}
  \label{rpa_sj}
 \end{center}
\end{figure}

\begin{figure}[tb]
 \begin{center}
  \includegraphics[clip,width=75mm]{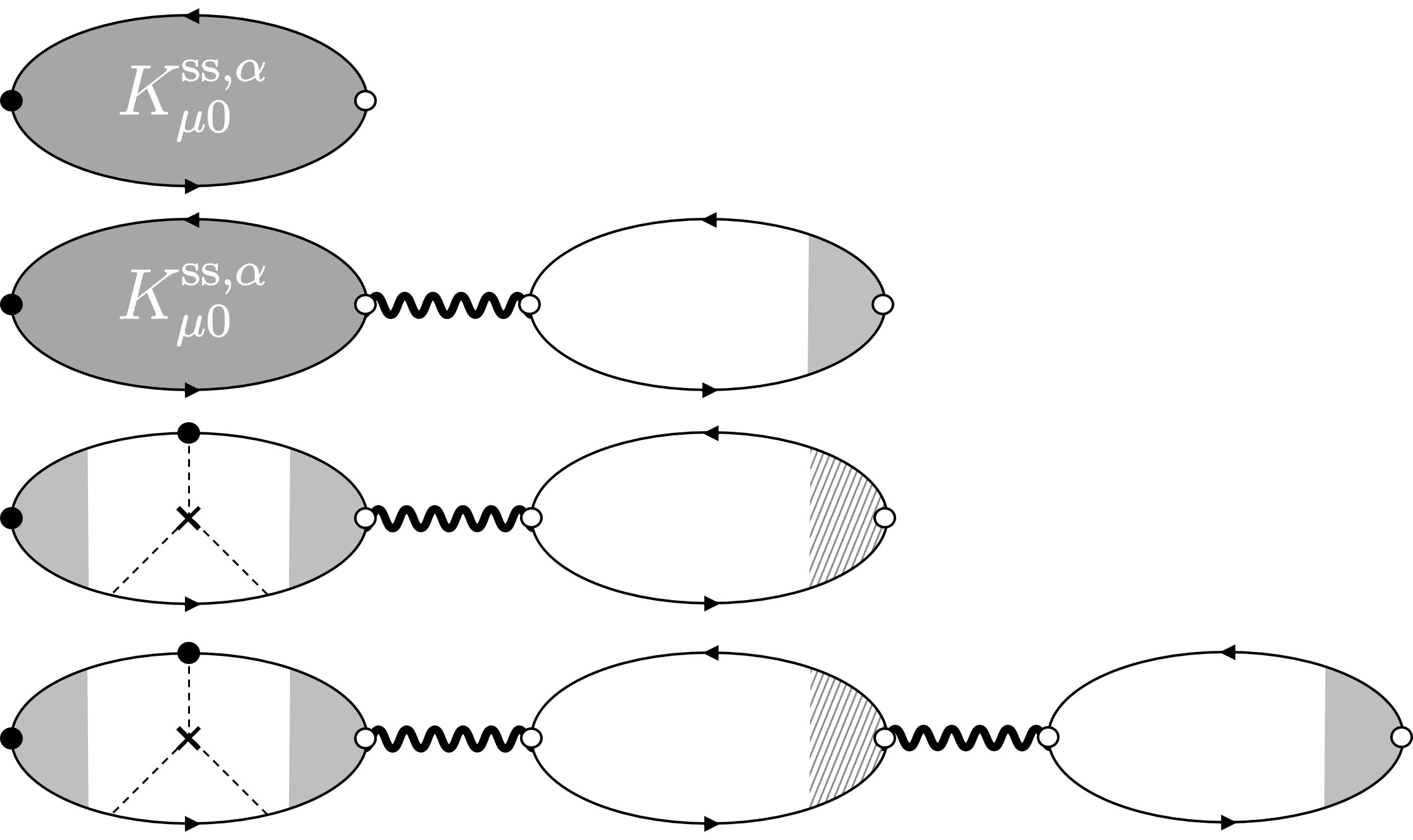}
  \caption{ 
 The skew-scattering contribution in the presence of long-range Coulomb interaction. 
  The leftmost gray ovals in the first and the second line are the response functions 
in the absence of the Coulomb interaction, given by Eqs.~(\ref{app_ss_00}) and (\ref{app_ss_i0}), 
respectively [see also Fig.~\ref{skew_mod}].
  The upside-down diagrams and the diagrams without ladder vertex corrections are also included in the calculation.}
  \label{rpa_ss}
 \end{center}
\end{figure}

 The spin-current response functions with side-jump and skew-scattering processes,
 including the RPA diagrams (see Figs.~\ref{rpa_sj} and \ref{rpa_ss}), are given by
\begin{widetext}
\begin{align}
 K^{\text{RPA},\alpha}_{\mu j} 
&= K^{\alpha}_{\mu j}
  - K^{\alpha}_{\mu 0} U^{\text{eff}}_{\bm q} \pi^{\rm cc}_{0j} (\bm q)  
  - X^{ \alpha}_{\mu 0} (\bm q+\bm Q) 
 \Bigl[ 1+\Lambda^{\text c}_0 (\bm q+\bm Q) \Bigr] \frac{U^{\text{eff}}_{\bm q+\bm Q}}{e^2}
 \Bigl[ K^{\text{cc}}_{0j} - K^{\text{cc}}_{00} U^{\text{eff}}_{\bm q} \pi^{\rm cc}_{0j} (\bm q) \Bigr]   ,
\end{align}
where $X^{\alpha}_{\mu \nu}=X^{\text{sj},\alpha}_{\mu \nu}+X^{\text{ss}, \alpha}_{\mu \nu}$ is the total response functions of SHE in the absense of the inhomogeneity $\delta n_{\text i}$. 
$K^{\alpha}_{\mu 0}$ describes spin accumulation and spin current in response to scalar potential 
given in Eqs.~(\ref{app_sj_00}), (\ref{app_sj_i0}), (\ref{app_ss_00}), and (\ref{app_ss_i0}).
 The spin accumulation and spin current are given by
\begin{align}
 \average{\hat \sigma^{\alpha}}^{\text{RPA(sj)}}_{\delta} 
&= \average{\hat \sigma^{\alpha}}^{\text{sj}}_{\delta},
\\
 \average{\hat j^{\alpha}_{\text s, i}}^{\text{RPA(sj)}}_{\delta} 
&= -\alpha^{\text{sj}}_{\text{SH}} \epsilon_{\alpha ij} Di(q_j +Q_j) 
      \frac{\average{\hat \rho_{\text e}}^{\text{RPA}}_{\delta}}{-e}
 - Di(q_i+Q_i) \average{\hat \sigma^{\alpha}}^{\text{sj}}_{\delta} - \delta D(\bm Q) iq_i 
                 \average{\hat \sigma^{\alpha}}^{\text{sj}}_0,
 \\
 \average{\hat \sigma^{\alpha}}^{\text{RPA(ss)}}_{\delta} 
&= - \frac{\alpha^{\text{ss}}_{\text{SH}}}{-e} 
       \frac{1}{D(\bm q+\bm Q)^2 - i\omega +\tau_{\text{sf}}^{-1}}
 \biggl\{ \Bigl[ i(\bm q+\bm Q) \times \average{\hat{\bm j}_{\text e}}^{\text{RPA}}_{\delta} \Bigr]_{\alpha}
  - \frac{\delta D \, \bm q \cdot (\bm q+\bm Q) 
   [i\bm q\times \average{\hat{\bm j}_{\text e}}_0]_{\alpha}}{Dq^2-i\omega + \tau^{-1}_{\text{sf}}}
\biggr\} ,
\\
 \average{\hat j^{\alpha}_{\text s, i}}^{\text{RPA(ss)}}_{\delta} 
&= \alpha^{\text{ss}}_{\text{SH}} \epsilon_{\alpha ij} \frac{\average{\hat j_{\text e, j}}^{\text{RPA}}_{\delta}}{-e} 
  - Di(q_i+Q_i) \average{\hat \sigma^{\alpha}}^{\text{RPA(ss)}}_{\delta} 
  - \delta D(\bm Q) iq_i \average{\hat \sigma^{\alpha}}^{\text{ss}}_0.
\end{align}
The results in the absense of the impurity inhomogeneity are given by
\begin{align}
\average{\hat \sigma^{\alpha}}_0^{\text{RPA}} &= \average{\hat \sigma^{\alpha}}_0 = - \alpha_{\text{SH}} \frac{\sigma_{\text c}}{-e} \frac{[i\bm q\times \bm E]_{\alpha}}{Dq^2-i\omega +\tau_{\text{sf}}^{-1}},
\\
\average{\hat j_{\text s ,i}^{\alpha}}_0^{\text{RPA}} &= \alpha_{\text{SH}} \epsilon_{\alpha ij} \frac{\average{\hat j_{\text e, j}}^{\text{RPA}}_0}{-e} - Diq_i \average{\hat \sigma^{\alpha}}_0  .
\end{align}
For calculation, we derive the response functions of spin accumulation and spin current to scalar potential as well.

\begin{align}
 K^{\text{sj},\alpha}_{00} &= 0  , 
\label{app_sj_00}
\\
 K^{\text{sj},\alpha}_{i0} 
&= \Bigl[ X^{\text{sj(b)},\alpha}_{i0}(\bm q+\bm Q) + X^{\text{sj(c)},\alpha}_{i0} (\bm q+\bm Q) \Bigr]
  \Pi^{\text{(R)}}_0(\bm Q) ,
\label{app_sj_i0}
\\
 K^{\text{ss},\alpha}_{00} 
&= \Bigl[ 1+\Lambda^{\text s}_0(\bm q+\bm Q) \Bigr] 
  \biggl[
  i\omega \alpha^{\text{ss}}_{\text{SH}} \frac{\delta n_{\text i}(\bm Q)}{n_{\text i}} \frac{\sigma_{\text c}}{-e} 
  \left( \frac{3\gamma}{v_{\text F}^2\pi N(\mu )} \right)^2
  \epsilon_{\alpha lm} I_l(\bm q+\bm Q) I_m (\bm q) + L^{\text{ss},\alpha}_{00} \biggr]
  \Bigl[ 1+\Lambda^{\text c}_0 (\bm q) \Bigr]  ,
\label{app_ss_00}
\\
 K^{\text{ss},\alpha}_{i0} 
&= \biggl[
   i\omega \alpha^{\text{ss}}_{\text{SH}} \frac{\delta n_{\text i}(\bm Q)}{n_{\text i}} 
    \frac{\sigma_{\text c}}{-e} \left( \frac{3\gamma}{v_{\text F}^2 \pi N(\mu )} \right) ^2
    \epsilon_{\alpha lm} I_m(\bm q)
   \Bigl[ I_{il} + \Lambda^{\text s}_i I_l  \Bigr]_{\bm q+\bm Q}
 +  L^{\text{ss}, \alpha}_{i0} + \Lambda^{\text s}_i (\bm q+\bm Q) L^{\text{ss},\alpha}_{00}   \biggr]
    \Bigl[ 1+\Lambda^{\text c}_0 (\bm q) \Bigr]
\nonumber \\
& \ \ \  + X^{\text{ss},\alpha}_{i0} (\bm q+\bm Q) \Pi^{\text{(R)}}_0 (\bm Q) .
\label{app_ss_i0}
\end{align}

\end{widetext}

\subsection{Results} 
\label{App:results_Coulomb}

 The results are given by formally the same expression as Eqs.~(\ref{cd_0_result})-(\ref{cc_PP}). 
 The only modification by the Coulomb interaction is in the charge diffusion propagator, 
$\Pi_{\rm c}({\bm q}) = [D(q^2+q_{\rm T}^2) - i\omega]^{-1}$ (for 3D) or 
$[Dq (q+q_{\rm T}) - i\omega]^{-1}$ (for 2D), 
from $ (Dq^2 - i\omega)^{-1}$. 
 The spin diffusion propagator remains the same, 
$\Pi_{\rm s}({\bm q}) = (Dq^2 - i\omega + \tau_{\text{sf}}^{-1})^{-1}$. 
 As a result, the effective field ${\bm F}$, defined by Eq.~(\ref{F_def1}), has the form, 
\begin{align}
 {\bm F} &=  {\bm E}  - Dq^2 \Pi_{\rm c}({\bm q}) {\bm E}_\parallel 
 - D q_{\rm T}^2 \Pi_{\rm c}({\bm q}) {\bm E}_\parallel , 
\label{F_Coulomb}
\end{align}
for 3D. 
 The second term ($\propto Dq^2$) describes the diffusion current as before. 
 The third term ($\propto Dq_{\rm T}^2$) represents a real electric field produced 
by charge accumulation, 
which is denoted by ${\bm E}^{\rm ind}_0$ in the text [Eq.~(\ref{eq:E^ind_0})].

\section{Screening of spin-orbit interaction}
\label{Breit}

 Starting from the Dirac equation that describes interacting two electrons, 
and then taking the weak-relativistic limit, 
Breit derived the relativistic corrections to the Coulomb (or electromagnetic) interaction.\cite{Breit}  
 Among many terms, let us focus on those relevant to SOI, 
\begin{align}
 {\cal H}_{\rm B} &= \frac{e^2 \hbar}{(2mc)^2} 
      \left[ \frac{{\bm r}_1 - {\bm r}_2}{|{\bm r}_1 - {\bm r}_2|^3} \times (2{\bm p}_2 - {\bm p}_1) \right] 
      \cdot {\bm \sigma}_1 
  + (1 \leftrightarrow 2)  , 
\label{eq:Breit}
\end{align}
where ${\bm p}_1 = -i\hbar \nabla_1$, ${\bm p}_2 = -i\hbar \nabla_2$, 
and the labels 1 and 2 specify each electron. 
 The first term ($\propto 2{\bm p}_2$) describes the Zeeman interaction of electron 1 
with the Amperian field produced by electron 2, 
and the second term ($\propto {\bm p}_1$) is the SOI for electron 1 in the electric field produced 
by electron 2. 
 Writing it in a second-quantized form, 
$H_{\rm B} = \frac{1}{2} \int d{\bm r}_1 \int d{\bm r}_2 \psi^\dagger ({\bm r}_1) \psi^\dagger ({\bm r}_2) 
                  {\cal H}_{\rm B} \psi ({\bm r}_2) \psi ({\bm r}_1)$, 
and making the Hartree approximation, 
the resulting one-body Hamiltonian reads 
\begin{align}
 {\cal H}_{\rm B}^{\rm Hartree}  
&=  - \frac{e^2 \hbar}{(2mc)^2} \int d {\bm r}_2 
      \left[ \frac{{\bm r}_1 - {\bm r}_2}{|{\bm r}_1 - {\bm r}_2|^3} \times  {\bm p}_1 \right] 
      \cdot {\bm \sigma}_1 \, 
    \delta n_{\rm el} ({\bm r}_2) 
\nonumber \\
&=   \frac{\hbar}{(2mc)^2}  
      (\nabla \delta V ({\bm r}_1) \times  {\bm p}_1 ) \cdot {\bm \sigma}_1 , 
\label{eq:Breit_SOI}
\end{align}
where 
$\delta V ({\bm r}_1) = e^2 \int d {\bm r}_2  \frac{\delta n_{\rm el} ({\bm r}_2)}{|{\bm r}_1 - {\bm r}_2|}$ 
is the change of the Hartree potential due to electron density modulation $\delta n_{\rm el} ({\bm r}_2)$. 
 This partly cancels the SOI field from the nucleus, and corresponds to the \lq\lq screening'' of the SOI. 
 In essence, this just means that the SOI is determined by the screened potential, 
namely, potential from the nucleus screened by electrons in the \lq\lq core'' region. 
 Therefore, the impurity SOI, given in the model, can be screened only when the electron density 
in the core region is changed, and this is very unlikely in ordinary situations in solid state physics.

\section{Integrals}
\label{int}

 The integrals $I_{\mu \nu}$ [Eq.~(\ref{app:I})] and $Y_{\mu \nu}$ [Eq.~(\ref{app:Y})] are calculated as
\begin{align}
 I_{0}(\bm q) &=I_{00}(\bm q) = \frac{\pi N(\mu )}{\gamma} \Bigl[ 1- \tau (Dq^2-i\omega ) \Bigr] ,
\label{i0}
\\
 I_{i}(\bm q) &=  I_{i0}(\bm q) = I_{0i}(\bm q) =  -Diq_i \frac{\pi N(\mu )}{\gamma} ,
 \\
 I_{ij} (\bm q) &= \delta_{ij}\frac{v_{\text F}^2}{3} \frac{\pi N(\mu )}{\gamma} 
                     \Bigl[ 1- \tau (Dq^2-i\omega ) \Bigr] , 
\end{align}
and 
\begin{align}
 Y_0(\bm q, \bm Q) &= \left( \frac{\pi N(\mu )}{\gamma } \right)^2 
                \Bigl\{ 1-\tau (Dq^2-i\omega ) 
\nonumber \\
& - \tau [D(\bm q+\bm Q)^2-i\omega ] -\tau D\bm q\cdot (\bm q+\bm Q) \Bigr\} ,
\label{y0}
\\
 Y_i(\bm q, \bm Q) &= -\left( \frac{\pi N(\mu )}{\gamma } \right)^2 Di ( 2q_i+Q_i ) ,
\\
 Y_{ij}(\bm q, \bm Q) 
&= \delta_{ij} \frac{v_{\text F}^2}{3}\left( \frac{\pi N(\mu )}{\gamma } \right)^2  
    \Bigl\{ 1-\tau (Dq^2-i\omega )
\nonumber \\
& - \tau [D(\bm q+\bm Q)^2-i\omega ] -\tau D\bm q \cdot (\bm q+\bm Q) \Bigr\} .
\end{align}

 The electromagnetic response functions [Eq.~(\ref{app:pi^cc})], 
evaluated with ladder vertex corrections, are given by 
\begin{align}
 \pi^{\text{cc}}_{00}(\bm q ) &= \frac{n_{\text e} \tau}{m} \frac{q^2}{Dq^2-i\omega},
\\
 \pi^{\text{cc}}_{i0} (\bm q ) 
&= \pi_{0i}^{\text{cc}} (\bm q ) = \frac{n_{\text e} \tau}{m} \frac{\omega q_i}{Dq^2-i\omega}.
\end{align}

\end{document}